\documentclass[pre, aps, floatfix, showpacs, preprint]{revtex4}
\usepackage{latexsym}
\usepackage{amssymb}
\usepackage{amsmath}
\usepackage{graphicx}
\usepackage{epsfig}
\usepackage{subfigure}

\newcommand{\stl}[1]{\mbox{$ \hspace{0.1em}
      \stackrel{\rule{0.4pt}{0.275ex}\hspace{0.40em} \!\!\!
      \overline{\hspace{0.06em}\vphantom{\rule{0.4pt}{0.0ex}}
      \hphantom{\mbox{$\displaystyle #1$}}
      \hspace{0.06em}  } \!\!\!\hspace{0.40em}\rule{0.4pt}{0.275ex}}
      {#1}\hspace{0.2em}$}}

\begin{document}
\title{Numerical Method of Lines for the Relaxational Dynamics of Nematic Liquid Crystals}
\author{A. K. Bhattacharjee}
\author{Gautam I. Menon}
\author{R. Adhikari}
\affiliation{The Institute of Mathematical Sciences,
C.I.T. Campus, Taramani, Chennai 600013, India}
\date{\today}

\begin{abstract}
We propose an efficient numerical scheme, based on the method
of lines, for solving the Landau-de Gennes equations describing the
relaxational dynamics of  nematic liquid crystals. Our method is 
computationally easy to implement, balancing requirements of efficiency 
and accuracy. We benchmark our method through 
the study of the following problems: the isotropic-nematic interface, 
growth of nematic droplets in the isotropic phase and the kinetics of 
coarsening following a quench into the nematic phase. Our results,
obtained through solutions of the full coarse-grained equations of 
motion with no approximations,
provide a stringent test of the de Gennes ansatz for the isotropic - nematic
interface, illustrate the anisotropic character of droplets in 
the nucleation regime and validate dynamical scaling in the coarsening 
regime.
\end{abstract}
\date{\today}
\maketitle

\section{\bf Introduction}
Liquid crystalline phases of matter are rich in examples of subtle order parameters, complex 
phase behaviour, and exotic defect structures \cite{degenpro}. In the nematic phase of 
liquid crystals, broken rotational symmetry produces elastic and hydrodynamic modes and topological 
defects of integer and half-integer charge \cite{pcl}. Understanding the statics and 
dynamics of nematics is difficult because the order parameter is neither a scalar or a vector, 
but a more complicated tensorial quantity constrained by symmetry and normalisation \cite{degenpro}. 
For example, the simplest relaxational dynamics which follows from a  time-dependent Ginzburg-Landau 
equation describing a nematic close to thermal equilibrium, entails the solution of a set of 5 coupled 
non-linear parabolic partial differential equations for the independent components of the order
parameter tensor ${\bf Q}$ \cite{degennes}. For all but the simplest situations, these equations 
do not have analytical solutions and thus need to be solved numerically. The efficient numerical 
computation of the solutions of the relaxational dynamics of nematics - nematodynamics - is the 
problem addressed here.

In this paper, we propose an efficient numerical scheme, based on the method of lines 
\cite{liskovets, leveque1}, for solving the Landau-de Gennes equations for nematodynamics. The method 
of lines (MOL) is a powerful technique for discretizing initial-value partial differential equations 
(PDEs). The essence of the MOL is semi-discretisation, where a PDE in spatial and temporal variables 
is discretised in the spatial variable only. This reduces the PDE to a system of ODEs in the temporal 
variable. The great advantage of the MOL is that there are powerful numerical methods implemented in 
general-purpose numerical libraries for the solution of systems of ODEs. The MOL also provides a great 
degree of freedom in discretising space, allowing one to chose from finite-difference, finite-volume, 
finite-element, spectral collocation, or any other suitable spatial discretisation. The strategy of 
semi-discretisation, where the spatial discretisation and temporal integration are treated as separate 
steps, allows for considerable flexibility, as both these operations can be optimised according to the demands 
of the problem to ensure maximum accuracy at minimum computational cost.

Here we combine a finite-difference spatial discretisation with a Runge-Kutta temporal integration to 
solve the relaxational equations of nematodynamics. We benchmark our method through the study of the 
following three problems: (a) the isotropic - nematic interface and tests of the de Gennes ansatz, 
(b) the kinetics of the growth of single nematic droplets in the isotropic phase and (c),  
coarsening kinetics following a quench from the isotropic into the nematic phase. To enable a comparison 
with previously published results our numerical results are in two dimensions, though the method and our 
code are applicable in one, two and three dimensions.

Our results are summarised as follows. Our numerical scheme permits a stringent test of the de Gennes 
ansatz for the uniaxial nematic isotropic interface \cite{degennes}. We show that biaxiality is absent 
in the interfacial region and that the hyperbolic tangent profile proposed by de Gennes provides an 
accurate fit to the numerical data. We show next that single nematic droplets placed in an isotropic 
solvent at isotropic-nematic coexistence coarsen anisotropically \cite{cutoedijks}, with initially 
circular droplets deforming into more ellipsoidal structures as time evolves. In our study of coarsening 
upon quenches from the isotropic phase, we observe the characteristic annihilation of integer and half 
integer charge defects (disclinations) as revealed by schlieren plots. For the coarsening problem, we 
find that order parameter correlation functions at different times can be rescaled to lie on a master 
curve, and that the coarsening exponent associated with the growing length scales is $1\over 2$. Our 
numerical results are in excellent agreement with analytical results where available and consistent 
with numerical results where such result exists. In addition, we extend previous results for each of 
the problems we study in significant ways.

In Section \ref{sec:2}, we summarise the order parameter description of nematics in 
terms of the symmetric, traceless order parameter tensor $Q_{\alpha \beta}$. We discuss the Landau - 
de Gennes free energy functional describing the free energy associated with order parameter configurations, 
the phase diagram for the free energy and the equations of relaxational dynamics. In section \ref{sec:3} 
we present the MOL scheme for partial differential equations and show how it may be applied to nematodynamics. 
In section \ref{sec:4} we discuss each of the applications described above in some detail, 
presenting our numerical results. We conclude by describing extensions of the present method to other problems 
in the dynamics of nematic liquid crystals  and compare our method with other methods available in the literature.

\section{Relaxational dynamics}
\label{sec:2}
Orientational order in the nematic phase is quantified through a second-rank, symmetric traceless tensor, 
$Q_{\alpha\beta}$ \cite{degenpro}. The principal axes of this tensor, obtained by diagonalizing ${\bf Q}$, 
specify the direction of ordering. The principal values represent the strength of ordering. Locally, nematic 
order is defined through 
${\bf Q}({\bf x},t) = \int du f({\bf x},{\bf u},t) {\stl{\bf uu}} \equiv \langle {\stl{\bf uu}}\rangle$ 
where $f({\bf x},{\bf u},t)$ is the molecular orientational distribution function at the position ${\bf x}$ 
at time $t$ which counts the number of molecules with long axis oriented in the direction ${\bf u}$, ${\stl X}$ 
defines the symmetric traceless part of an arbitrary real second rank tensor ${\bf X}$ with 
\begin{equation}
{\stl X} = \frac{1}{2}\left(X + X^{T}\right) - \frac{1}{d} Tr(X),
\end{equation}
where $X^{T}$ denotes the transpose of the tensor i.e. $(X^{T})_{\alpha \beta} = X_{\beta \alpha}$
and $d$ denotes the dimension. The average $<\cdot> \equiv \int du f({\bf x}, {\bf u}, t)(\cdot)$.

The three principal axes specify the director ${\bf n}$, the codirector ${\bf l}$ and
the joint normal to these, ${\bf m}$. Given the two principal values $S$ and $T$, the 
order parameter can be written as
\begin{equation}
\label{Qtensor}
Q_{\alpha \beta}=\frac{3}{2}S(n_{\alpha}n_{\beta} - \frac{1}{3}\delta_{\alpha \beta}) + 
\frac{1}{2}T(l_{\alpha}l_{\beta} - m_{\alpha}m_{\beta}),
\end{equation}
where $\alpha, \beta \equiv x, y, z$ and,
\begin{eqnarray}
S &=& \langle {P_{2}(\cos \theta)}\rangle = \langle {(\cos^{2}\theta - \frac{1}{3})}\rangle, \\
T &=& \langle {\sin^{2}\theta \cos2\phi}\rangle,
\end{eqnarray}
are measures of alignment with $-\frac{1}{3} \leq S \leq \frac{2}{3}$, $0 \leq T < 3S$.
$S = \frac{2}{3}$ corresponds to the nematic phase whereas $S = 0$ to the isotropic phase. 
The (polar) angle between {\bf n} and {\bf u} is  $\theta$, while the (azimuthal) 
angle between {\bf l} and {\bf u} is $\phi$.

In the uniaxial nematic phase, $T=0$ and the tensor ($\ref{Qtensor}$) takes the form 
${\bf Q} = (3/2)S{\stl{\bf nn}}$. In the biaxial phase ${\bf Q} = (3/2)S{\stl { \bf nn} } + (1/2)T({\bf ll} -
{\bf mm})$. The tensor {\bf Q} is diagonal in a coordinate system aligned with the principal axes,
\begin{eqnarray}
\label{Qdiagmatrixform}
{\bf Q} = \left( \begin{array}{ccc} -(S + T)/2 &  0 & 0 \\ 0 &
-(S - T)/2 &  0\\ 0 &  0& S \end{array}\right).
\end{eqnarray}

In a frame of reference which is not aligned with the principal axes, ${\bf Q}$ must be expanded in a set of 
basis tensors $T^{i}_{\alpha\beta}$ \cite{reinhesskrog} which are symmetric, traceless and orthonormal, 
\begin{equation}
\label{orthbasis}
Q_{\alpha\beta}({\bf x}, t) = \sum_{i=1}^{5}a_{i}({\bf x}, t)T^{i}_{\alpha\beta}.
\end{equation}
Explicitly, these are ${\bf T}^{1} = \sqrt{\frac{3}{2}} \stl{{\bf \hat{z}\hat{z}}},
{\bf T}^{2} =  \sqrt{\frac{1}{2}} ({\bf \hat{x} \; \hat{x} - \hat{y} \; \hat{y}}),
{\bf T}^{3} = \sqrt{2}\; \stl{{\bf \hat{x} \; \hat{y}}},
{\bf T}^{4} = \sqrt{2}\; \stl{{\bf \hat{x} \; \hat{z}}}$ 
${\bf T}^{5} = \sqrt{2}\; \stl{{\bf \hat{y} \; \hat{z}}}$.

The Landau-Ginzburg free energy functional $F$ \cite{degennes} is obtained from a local expansion in
powers of rotationally invariant combinations of the order parameter ${\bf Q}({\bf x},t)$,
\begin{equation}
\label{localfrener}
\mathcal{F}_{h}[{\bf Q}] = \frac{1}{2}ATr{\bf Q}^{2}  +
\frac{1}{3}BTr{\bf Q}^{3} + \frac{1}{4}C(Tr{\bf Q}^{2})^{2} +
E^{\prime}(Tr{\bf Q}^{3})^{2}.
\end{equation}
To this, non-local terms arising from rotationally invariant combinations of gradients of
the order parameter must be added \cite{degennes}:
\begin{equation}
\mathcal{F}_{el}[{\bf \partial Q}] = \frac{1}{2}L_{1}(\partial_{\alpha}
Q_{\beta\gamma})(\partial_{\alpha}Q_{\beta\gamma}) +
\frac{1}{2}L_{2}(\partial_{\alpha}Q_{\alpha\beta})(\partial_{
\gamma}Q_{\beta\gamma}),
\end{equation}
where $\alpha, \beta, \gamma$ denote the Cartesian directions in the local
frame, and $L_{1}$ and $L_{2}$ are elastic constants. Surface terms of the
same order in gradients may also be added, but we omit them in the present work.

In the local free energy density Eq.($\ref{localfrener}$), $A = A_{0}(1 - T/T^{*})$, 
where $T^{*}$ denotes the supercooling transition temperature. From the inequality
$\frac{1}{6}(Tr{\bf Q}^{2})^{3} \geq (Tr {\bf Q}^{3})^{2}$, higher powers of $Tr{\bf Q}^{3}$ 
can be excluded for the description of the uniaxial phase. Thus the uniaxial case is 
described by $E^{\prime}$ = 0 whereas $E^{\prime} \neq 0$ for the biaxial phase. 
For nematic rod-like molecules $B<0$ whereas for disc-like molecules, $B>0$. The 
quantities C and $E^{\prime}$ are always taken to be positive to ensure 
stability and boundedness of the free energy in both the isotropic and nematic phases.
The total free energy including local and gradient terms is 
\begin{widetext}
\begin{equation}
\label{frener}
F = \int d^3{\bf x} [\frac{1}{2}ATr{\bf Q}^{2} + \frac{1}{3}BTr{\bf Q}^{3} + 
\frac{1}{4}C(Tr{\bf Q}^{2})^{2} + E^{\prime}(Tr{\bf Q}^{3})^{2} + 
\frac{1}{2}L_{1}(\partial_{\alpha}Q_{\beta\gamma})(\partial_{\alpha}Q_{\beta\gamma}) 
+ \frac{1}{2}L_{2}(\partial_{\alpha}
Q_{\alpha\beta})(\partial_{\gamma}Q_{\beta\gamma})].
\end{equation}
\end{widetext}

The first order isotropic to uniaxial nematic transition at the critical value $S = S_{c}$ 
is obtained from the equations,
\begin{eqnarray}
A &=& \frac{3}{4}CS_c^{2} + \frac{9}{4}E^{\prime}S_c^{4}, \\
B &=& -\frac{9}{2}CS_c - 9E^{\prime}S_c^{3}.
\end{eqnarray}
The second order uniaxial to biaxial transition at the critical value $S = S^{\prime}_c$ is 
obtained from,
\begin{eqnarray}
A &=& - \frac{9}{2}C{S^{\prime}_c}^{2}, \\
B &=& - \frac{45}{4}E^{\prime}{S^{\prime}_c}^{3}.
\end{eqnarray}

\begin{figure}
\centering
\includegraphics[width=3.8in, height=3in]{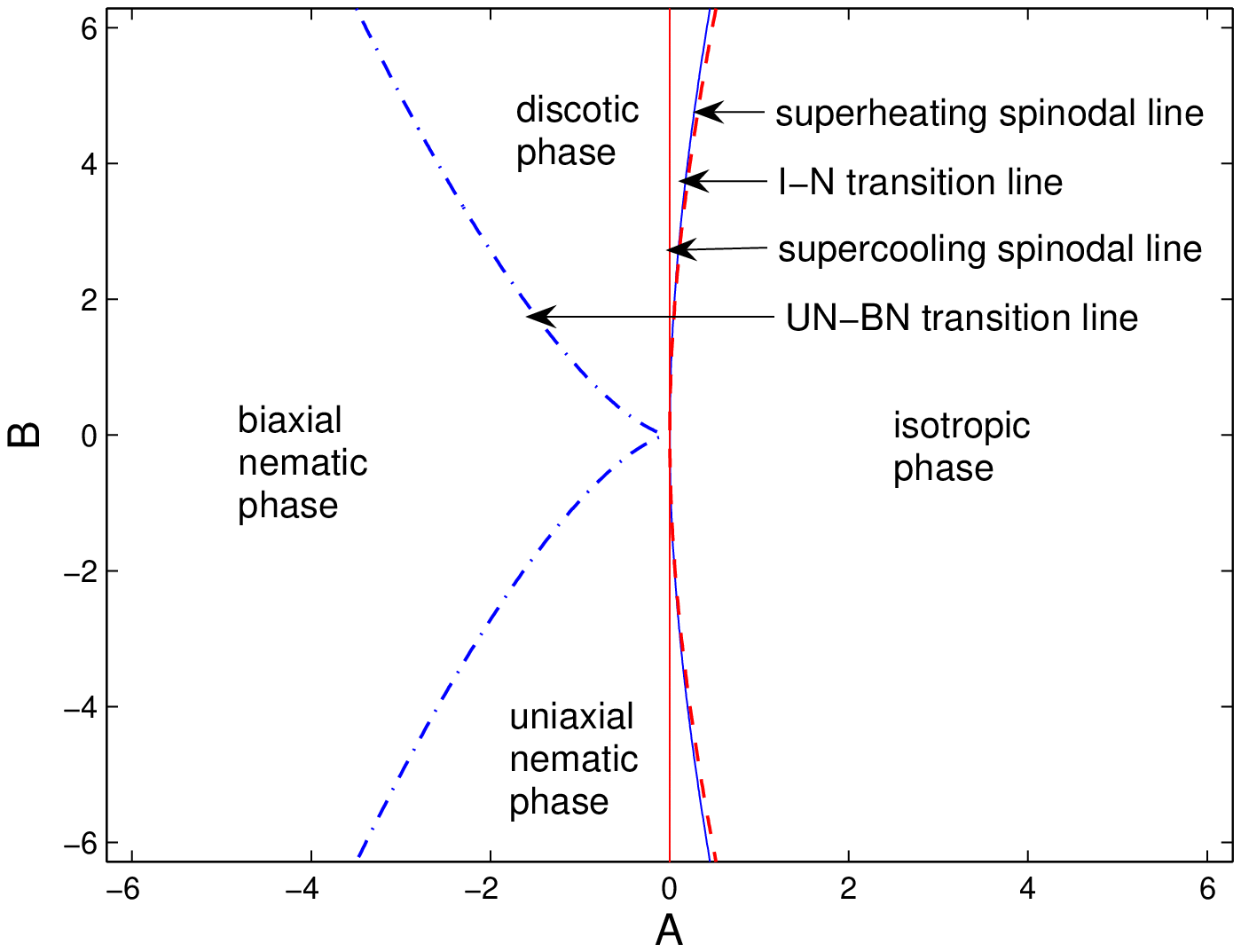}
\caption{(Color online) The mean field phase diagram obtained from the Landau-de Gennes 
free energy ($\ref{localfrener}$). The phase boundaries are given by solutions of the 
following algebraic equations : $4B^{4}E + B^{2}C(3C^{2} - 144AE^{\prime}) =
A(81C^{4} - 864AC^{2}E^{\prime} + 2304A^{2}E^{\prime^{2}})$ for the isotropic to uniaxial 
nematic transition; $25A^{3}E^{\prime^{2}} = -18B^{2}C^{3}$ for the uniaxial to biaxial 
nematic transition; A=0 for the supercooling spinodal; $9B^{4}E^{\prime} + 8B^{2}(C^{3} -
36ACE^{\prime}) = 192A(C^{2} - 4AE^{\prime})^{2}$ for the superheating spinodal.}
\label{fig:phasediag}
\end{figure}

In the phase diagram of Fig.~(\ref{fig:phasediag}), there are lines separating the nematic 
phase with $S > 0$ from the discotic phase in addition to the isotropic to nematic transition 
lines. There are also two uniaxial to biaxial nematic second order lines as well as the two 
first order isotropic-uniaxial nematic lines. One of these transition lines
marks the transition to the nematic 
phase with $S > 0$ and the other for the discotic phase with $S < 0$. The four phases meet 
at the bicritical Landau point $A = B = 0$. At the bicritical point, the isotropic to nematic 
transition is second order rather than first order. 

In the absence of thermal fluctuations and hydrodynamic flow, the equation of motion of the 
order parameter can be taken to be relaxational \cite{degennes}, 
\begin{equation}
\label{Qdynamics}
\partial_{t}Q_{\alpha\beta}({\bf x}, t) = - \Gamma_{\alpha
\beta\mu\nu} {\delta F\over\delta Q_{\mu\nu}},
\end{equation}
with
\begin{equation}
\Gamma_{\alpha\beta\mu\nu} = \Gamma[\delta_{\alpha\mu}\delta_{\beta\nu} +
\delta_{\alpha\nu}\delta_{\beta\mu} - \frac{2}{3}\delta_{\alpha\beta}
\delta_{\mu\nu}].
\end{equation}
The kinetic coefficient $\Gamma_{\alpha\beta\mu\nu}$ ensures that the dynamics preserves the 
symmetry and tracelessnes of the order parameter. In the above, $\Gamma$ is a constant. 

With the choice of the Landau-de Gennes free energy, the equation of motion takes the form
\begin{eqnarray}
\label{Qequation}
\partial_{t}Q_{\alpha\beta}({\bf x}, t) &=& - \Gamma \;[(A +
C TrQ^{2})Q_{\alpha\beta}({\bf x}, t) + (B + 6E^{\prime}TrQ^{3})
\stl{Q^{2}_{\alpha\beta}({\bf x}, t)} \nonumber \\
&& - L_{1}\nabla^{2}Q_{\alpha\beta}({\bf x}, t) -
L_{2}\stl{\nabla_{\alpha}(\nabla_{\gamma}Q_{\beta\gamma}({\bf x}, t))}].
\end{eqnarray}
The equations can be projected onto the tensorial basis $T^{i}_{\alpha\beta}$ to give a set of equations for the independent components $a_i$,
\begin{equation}
\label{dynamiceqn1}
\partial_{t}a_{i} = - \Gamma \;[(A + C TrQ^{2})a_{i} + (B +
6E^{\prime}TrQ^{3})T_{\alpha\beta}^{i}\stl{{Q_{\alpha\beta
}^{2}}} - L_{1}\nabla^{2}a_{i} - L_{2} \stl{T^{i}_{\alpha\beta}T^{j}_{\beta\gamma}
\partial_{\alpha}\partial_{\gamma}a_{j}}].
\end{equation}
Using the orthonormality of the basis tensors $T_{\alpha\beta}^{i} T_{\beta\alpha}^{j} = \delta_{ij}$, this can be 
expanded to the set of five equations,
\begin{widetext}
\begin{eqnarray}
\partial_{t}a_{1} &=& - \Gamma \;[(A + C TrQ^{2})a_{1} + (B +
6E^{\prime}TrQ^{3})\frac{1}{\sqrt{6}}(a_{1}^{2} - a_{2}^{2} -
a_{3}^{2} + \frac{a_{4}^{2}}{2} + \frac{a_{5}^{2}}{2}) - \nonumber \\
&& L_{1}\nabla^{2}a_{1} - L_{2}(\frac{1}{6}\nabla^{2}a_{1} +
\frac{1}{2}\partial^{2}_{z}a_{1} + \frac{1}{\sqrt{12}}((
\partial^{2}_{y} - \partial^{2}_{x})a_{2} + \partial_{x}\partial_{z}
a_{4} + \partial_{y}\partial_{z}a_{5}) - \nonumber \\
&& \frac{1}{\sqrt{3}}\partial_{x}\partial_{y}a_{3})], \\
\partial_{t}a_{2} &=& - \Gamma \;[(A + C TrQ^{2})a_{2} + (B +
6E^{\prime}TrQ^{3})(-\sqrt{\frac{2}{3}}a_{1}a_{2} + \frac{a_{4}^{2}}
{\sqrt{8}} - \frac{a_{5}^{2}}{\sqrt{8}}) - \nonumber \\
&& L_{1}\nabla^{2}a_{2} - L_{2}(\frac{1}{\sqrt{12}}(\partial^{2}_{y} -
\partial^{2}_{x})a_{1} + \frac{1}{2}((\partial^{2}_{x} +
\partial^{2}_{y})a_{2} + \partial_{x}\partial_{z}a_{4} -
\partial_{y}\partial_{z}a_{5}))], \\
\partial_{t}a_{3} &=& - \Gamma \;[(A + C TrQ^{2})a_{3} + (B +
6E^{\prime}TrQ^{3})(-\frac{2a_{1}a_{3}}{\sqrt {6}} +
\frac{a_{4}a_{5}}{\sqrt {2}}) - \nonumber \\
&& L_{1}\nabla^{2}a_{3} - L_{2}(-\frac{1}{\sqrt{3}}\partial_{x}
\partial_{y}a_{1} + \frac{1}{2}((\partial^{2}_{x} + \partial^{2}_{y})
a_{3} + \partial_{y}\partial_{z}a_{4} + \partial_{x}\partial_{z}a_{5}))], \\
\partial_{t}a_{4} &=& - \Gamma \;[(A + C TrQ^{2})a_{4} + (B +
6E^{\prime}TrQ^{3})(\frac{a_{1}a_{4}}{\sqrt {6}} + \frac{a_{2}a_{4}}
{\sqrt {2}} + \frac{a_{3}a_{5}}{\sqrt {2}}) - \nonumber \\
&& L_{1}\nabla^{2}a_{4} - L_{2}(\frac{1}{\sqrt{12}}\partial_{x}
\partial_{z}a_{1} + \frac{1}{2}((\partial^{2}_{x} + \partial^{2}_{z})
a_{4} + \partial_{x}\partial_{z}a_{2} + \partial_{y}\partial_{z}a_{3} +
\nonumber \\
&& \partial_{x}\partial_{y}a_{5}))], \\
\label{dynamiceqn2}
\partial_{t}a_{5} &=& - \Gamma \;[(A + C TrQ^{2})a_{5} + (B +
6E^{\prime}TrQ^{3})(\frac{a_{1}a_{5}}{\sqrt {6}} + \frac{a_{3}a_{4}}
{\sqrt {2}} - \frac{a_{2}a_{5}}{\sqrt {2}}) - \nonumber \\
&& L_{1}\nabla^{2}a_{5} - L_{2}(\frac{1}{\sqrt{12}}\partial_{y}
\partial_{z}a_{1} + \frac{1}{2}((\partial^{2}_{y} + \partial^{2}_{z})
a_{5} - \partial_{y}\partial_{z}a_{2} + \partial_{x}\partial_{z}a_{3} +
\nonumber \\
&& \partial_{x}\partial_{y}a_{4}))].
\end{eqnarray}
\end{widetext}
Note that these equations are parabolic, non-linear, contain anisotropic 
gradient terms and are coupled strongly to each other. The efficient 
computation of the solutions to these equations in a variety of physical 
situations is described in succeeding sections of this paper.

\section{Method of Lines discretization}
\label{sec:3}
The equations for the relaxational dynamics of the orientational order parameter presented in the 
previous section are a set of five coupled non - linear parabolic differential equations. 
No analytical solutions are available for these equations in general and 
efficient and accurate numerical methods must therefore be sought. Below we present such a method, based 
on the strategy of semi-discretisation, whereby an initial value PDE is discretised only in the 
spatial variables to yield a set of coupled ODEs. These ODE's  can be solved using powerful general 
purpose solvers. 

In the literature, this semi-discretisation strategy goes by the name of the `method of lines' (MOL)
\cite{liskovets, leveque1}, since the solutions are obtained along fixed lines in the space - time plane. 
To illustrate this, consider a parabolic initial-value problem in a single spatial variable $x$, the 
diffusion equation for a scalar $\psi(x, t)$:
\begin{equation}
\partial_t \psi(x, t) = D\nabla^2\psi(x, t).
\end{equation}
The MOL discretisation proceeds by restricting the field $\psi(x, t)$ to a set of $N$ discrete 
`collocation' points, $x_n = n\Delta x$,  $n = 1, 2, \ldots, N$, spaced $\Delta x$ apart, and then uses 
a discrete approximation for the Laplacian based on these points. The simplest of these approximations 
is based on local polynomial interpolation resulting in finite difference schemes \cite{leveque1}. 
However, the MOL is not restricted to finite differences. When high accuracy is needed, global interpolation 
based on trigonometric or Chebyshev polynomials can be used to generate spectral approximations of 
derivatives \cite{trefethen}. When conservation laws need to be respected, finite volume approximations 
to the derivatives can be used \cite{leveque2}. In complicated geometries, a finite-element discretisation 
may be the most appropriate. In this example, the simplest approximation is based on nearest-neighbour finite 
differences, for which
\begin{equation}
(\nabla^2\psi)(x_n)={1\over (\Delta x)^2}(\psi(x_{n+1}) - 2\psi(x_n) + 
\psi(x_{n-1})) + O((\Delta x)^2).
\end{equation}
Inserting this into the diffusion equation, we obtain a set of coupled 
ordinary differential equations
\begin{equation}
\partial_t \psi(x_n, t) = {1\over (\Delta x)^2}(\psi(x_{n+1}) - 
2\psi(x_n) + \psi(x_{n-1})).
\end{equation}
This coupled set of ordinary differential equations, together with initial 
and boundary conditions, can now be integrated with a suitable numerical 
integration scheme which ensures accuracy, efficiency, and stability. 
It is at this stage that the flexibility of the MOL is most apparent, 
since any number of numerical integration schemes can be implemented 
without affecting the spatial discretisation. Depending on the nature 
of the system of ODEs, the optimal choice may be either an explicit scheme, an
 implicit scheme, or one which is designed to  handle stiffness. In 
the above example, for instance, it is well known that an explicit Euler 
integration scheme leads to an instability unless the time step $\Delta t$ 
is constrained by the Courant-Friedrichs-Lewy condition 
$\Delta t < (\Delta x)^2 / 2 D$ \cite{leveque1}. In contrast, an implicit integration 
scheme based on the trapezoidal rule gives the stable Crank-Nicolson update \cite{leveque1}. 
In general, semi-discretisation followed by numerical quadrature provides an 
elegant way of deriving many of the well-known schemes for parabolic PDEs. 

The practical implementation of PDE solvers using the MOL discretisation 
is simple since the spatially discretised ODE system can be passed directly 
to a general purpose ODE solver. The complexity, both algorithmic and computational, 
 in the numerical integration can thereby be transferred directly to the ODE solver library.

We have followed the MOL discretisation to construct a solver for PDEs 
describing the dynamics of the orientational order parameter presented 
in the previous section. We have used standard nearest-neighbour 
second-order accurate finite difference formulae for first and second 
derivatives in the spatial discretisation \cite{abramstegun}. More accurate difference 
approximations to derivatives can be obtained using Fornberg's general 
formula \cite{fornberg}. For the temporal integration, we have used the ODE solver routines 
in the GNU Scientific Library. For the problems we study here, we find that 
an explicit multi-stage integrator gives a good compromise between accuracy, 
stability and computation expense. 

Finally, it should be mentioned that the MOL discretisation is not restricted to parabolic initial 
value problems, but also applies to hyperbolic and mixed parabolic - hyperbolic PDEs \cite{liskovets}. 
This allows the method described in this paper to be extended to situations where effects of 
advection from hydrodynamic flow need to be accounted for.

\section{Applications}
\label{sec:4}
In this section we report benchmarking results using the methodology described 
in the previous section for the following experimentally relevant situations: the 
properties of the isotropic-nematic interface at coexistence, the kinetics of droplet
evolution between the binodal and spinodal lines
and spinodal decomposition into the nematic phase. In each case, we compare 
our numerical results with available analytical and numerical results in the 
literature.    

A suitable non-dimensionalisation of the equations of motion before they are discretised is essential for controlling
artefacts and error introduced by the discretisation. For our problem, we non-dimensionalise the governing equations 
(\ref{Qequation}) using the values of dimensionful quantities at coexistence ($T=T_c$). The resulting dimensionless 
quantities are superscripted with an asterisk, while the dimensionful quantities are subscripted with \textit{c}. We 
non-dimensionalise the order parameter $Q^{*}_{\alpha\beta} = Q_{\alpha\beta}/S_{c}$ by the value of the strength of 
ordering at coexistence $S_c =-2B/9C$. The non-dimensionalised strength of ordering is then $S^{*} = S/S_c = 3(1 + 
\sqrt{1 - 24AC/B^2})/4$. The gradient terms in the free energy are non-dimensionalised by the length $l_c = 
\sqrt{54C(L_{1} + 2L_{2}/3)/B^{2}}$, which is related to the width of the isotropic-nematic interface at 
coexistence. The free energy is non-dimensionalised as $F^{*} = F/F_{c}$, where $F_{c} = 9CS_{c}^{4}/16$ is 
proportional to the free energy barrier between the isotropic and nematic minima at coexistence. Finally, 
time is non-dimensionalised by the characteristic relaxation time $\tau = (\Gamma F_c)^{-1}$. 

To solve the equations of motion Eq.(\ref{dynamiceqn1}), we choose the discretisation length 
$\Delta x = \Delta y = 1$ and the integration time-step $\Delta t = 1$. These define simulation units of 
length and time. To ensure that discretisation errors and artefacts are kept to a minimum, the discretisation 
length must be much smaller than the characteristic length $\Delta x = \Delta y \ll l_c$.  The integration time 
step must be much smaller than the characteristic time scale $\Delta t \ll \tau$. The dimensionless 
discretisation scales must then satisfy $\Delta x^{*}  = \Delta x /l_c \ll 1$, 
$\Delta y^{*}  = \Delta y/l_c \ll 1$ and $\Delta t^{*} = \Delta t/\tau \ll 1$. Our simulations 
are performed maintaining the above conditions, on grid sizes ranging from $32 \times 32$ to $256 \times 256$, 
with periodic boundary conditions, using  the 4th-order Runge - Kutta method as implemented in the GNU 
Scientific Library for the ODE solver.

\subsection{The isotropic-nematic interface at coexistence}

The isotropic-nematic interface was studied in a seminal paper \cite{degennes} by de Gennes within the framework 
of a Landau-Ginzburg description. To render the problem analytically tractable, de Gennes made a 
specific assumption regarding the variation of the components of the order parameter across the 
interface. For an infinitely extended interface where, by homogeneity, variations perpendicular 
to the interface alone are allowed, de Gennes assumed that the only quantity which changed across 
the interface was the strength of ordering $S$. In the de Gennes ansatz, there is no biaxiality 
and no variation of the director across the interface. This reduces a problem with five 
degrees of freedom to a more manageable problem involving only a single degree of freedom. The 
variation of the ordering strength $S$ along the  coordinate $z$ normal to the interface located 
at $z_0$ can then be obtained analytically as,
\begin{equation}
\label{ansatzS}
S(z) = {S_c\over 2}(1-\tanh\frac{z-z_0}{w}),
\end{equation}
where $w = \sqrt{2}/S^{*}$ is the non-dimensional interfacial width.

We have verified this remarkable ansatz with a direct numerical solution. In
our numerical calculations, a strip of nematic interface was sandwiched between 
two isotropic domains with periodic boundary conditions. The system was
then allowed to relax to the minimum of the free energy. The parameters were
chosen such that the width $w \gg \Delta z = 1$, ensuring that discretisation errors were kept to a minimum.

\begin{figure}
   \centering
    \includegraphics[width=4.0in, height=3.5in, angle=0]{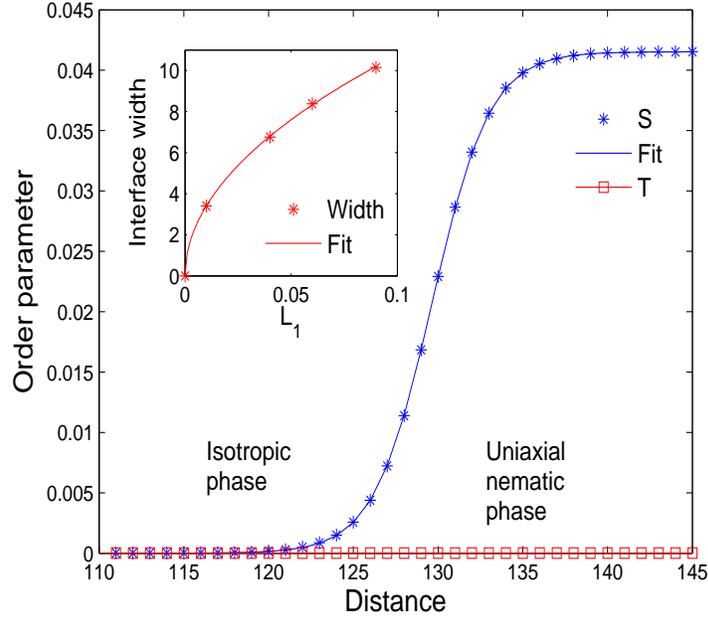}
\caption{(Color online) Variation of the degree of alignment (S) and biaxiality 
(T) across the nematic - isotropic interface with planar anchoring. Symbols 
represent numerical data, solid curves are the de Gennes ansatz (\ref{ansatzS}). 
The numerical parameters are A = 0.0035, B = -0.5, C = 2.67, $E^{\prime} = 0, 
L_{1} = 0.01, L_{2} = 0, \Gamma = 1/20$, grid size $8 \times 512$ and nematic 
strip width = $256$. The inset shows the variation of the interfacial width with the elastic 
constant $L_1$. The expected quadratic variation is accurately reproduced.}
\label{fig:ST}
\end{figure}

The resulting profiles for the variation of $S$ and $T$ are shown in 
Fig.~(\ref{fig:ST}). The values obtained for $T$ are consistent with 
de Gennes' assumption of vanishing biaxiality.  The variation of $S$ at 
each of the two isotropic-nematic interfaces were fitted, using the 
least-squares method, to the analytical profile, with saturation value of 
the order $S_c$, the location of the interface $z_0$ and the interface 
width $w$ as fitting parameters. As shown in the inset to Fig.~(\ref{fig:ST}),  
fitted values of $w$ agree remarkably well with the analytic result for a range
of parameter values. The agreement is accurate to within a fraction of a percent. 
Similar results were obtained for the saturation value of the order parameter.
This benchmark clearly demonstrates the accuracy of the MOL scheme in reproducing the 
equilibrium limit of Eqn.(\ref{Qdynamics}).

The de Gennes ansatz also predicts that the energies of planar and 
homeotropic anchoring are identical when elasticity is isotropic 
($L_2 = 0$). We compared the value of the free energies of the interface 
for both planar and homeotropic anchoring of the director. To machine precision, 
these answers are identical. Our results for this problem represent the first direct 
verification of the de Gennes ansatz retaining all degrees of freedom of the orientational tensor.

\subsection{Nematic droplet in an isotropic background}

Since the isotropic-nematic transition is first order, there exists 
a regime of parameters  where the kinetics proceeds by nucleation. 
In the phase diagram of Fig.~(\ref{fig:phasediag}) this regime is 
bounded by the binodal and spinodal lines. A droplet of nematic phase 
in an isotropic background shrinks if it is smaller than a 
critical size $R_c$. Droplets which are larger than $R_c$ 
grow in size till they expand to the size of the system. A droplet of 
size $R_c$ is thus metastable.

The development in time of a fluctuation-induced droplet is a key 
factor in determining the nucleation rate \cite{cutoedijks}. Unlike conventional
isotropic fluids, the nucleus in a nematic is not expected to be 
spherical \cite{bernfank}. Allowing the shape to deviate from perfect sphericity 
reduces the total energy, which is a sum of the elastic energy 
associated with the director deformation in the bulk and a surface 
energy associated with the anchoring condition at the interface of 
the droplet. Thus, determining the droplet shape of least energy involves
a minimisation over the strength of ordering as well as the director degrees
of freedom \cite{herring, chandrasekhar, virga}.

In our simulations we have studied the time evolution of droplet 
shapes. For studying conformational changes, our initial condition
was chosen to be  $S = S_{c}$, with the  director anchored at an 
angle of $\pi/4$ with the nematic - isotropic interface inside the 
droplet. We took $S = 0$ in the isotropic region. We relaxed the system at
a temperature intermediate between the binodal and spinodal temperatures.

\begin{figure}
  \subfigure[$\; t = 0 $]{\label{fig:L2zt0}
    \centering
 \includegraphics[width=1.5in, height=1.2in, angle=0]{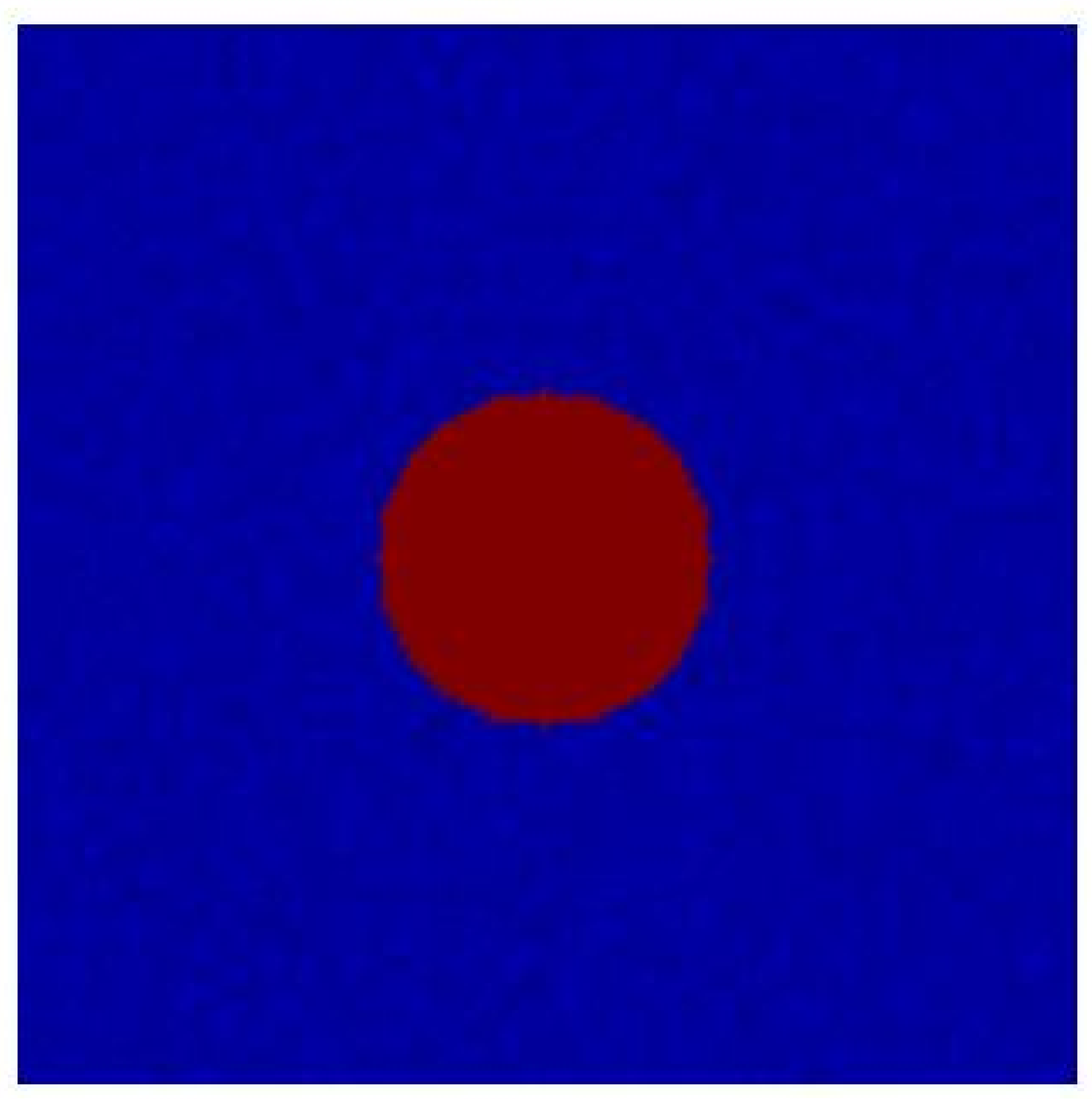}
}
  \subfigure[$\; t = 1000 $]{\label{fig:L2zt1k}
    \centering
 \includegraphics[width=1.5in, height=1.2in, angle=0]{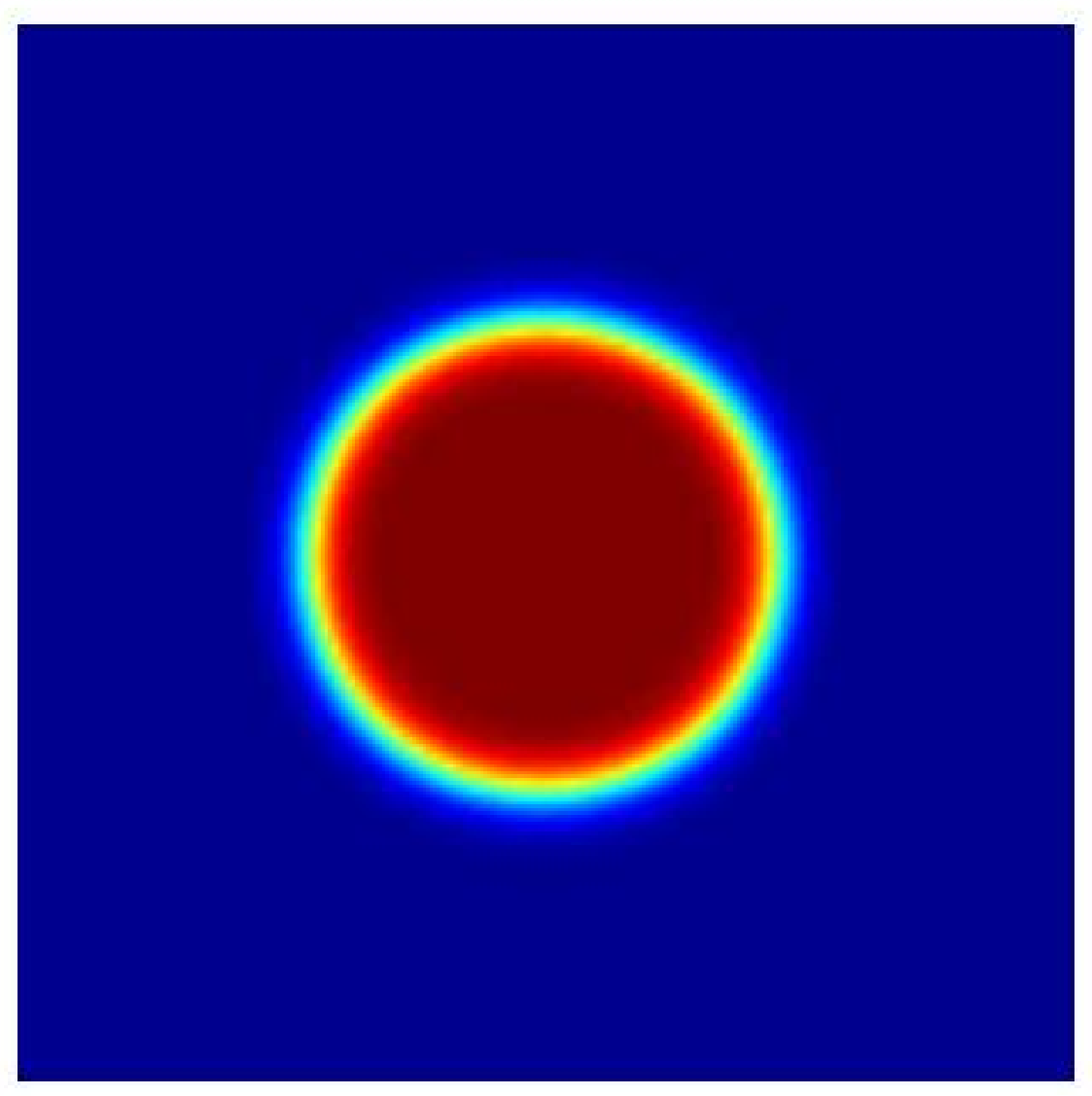}
}
  \subfigure[$\; t = 2000 $]{\label{fig:L2zt2k}
    \centering
 \includegraphics[width=1.5in, height=1.2in, angle=0]{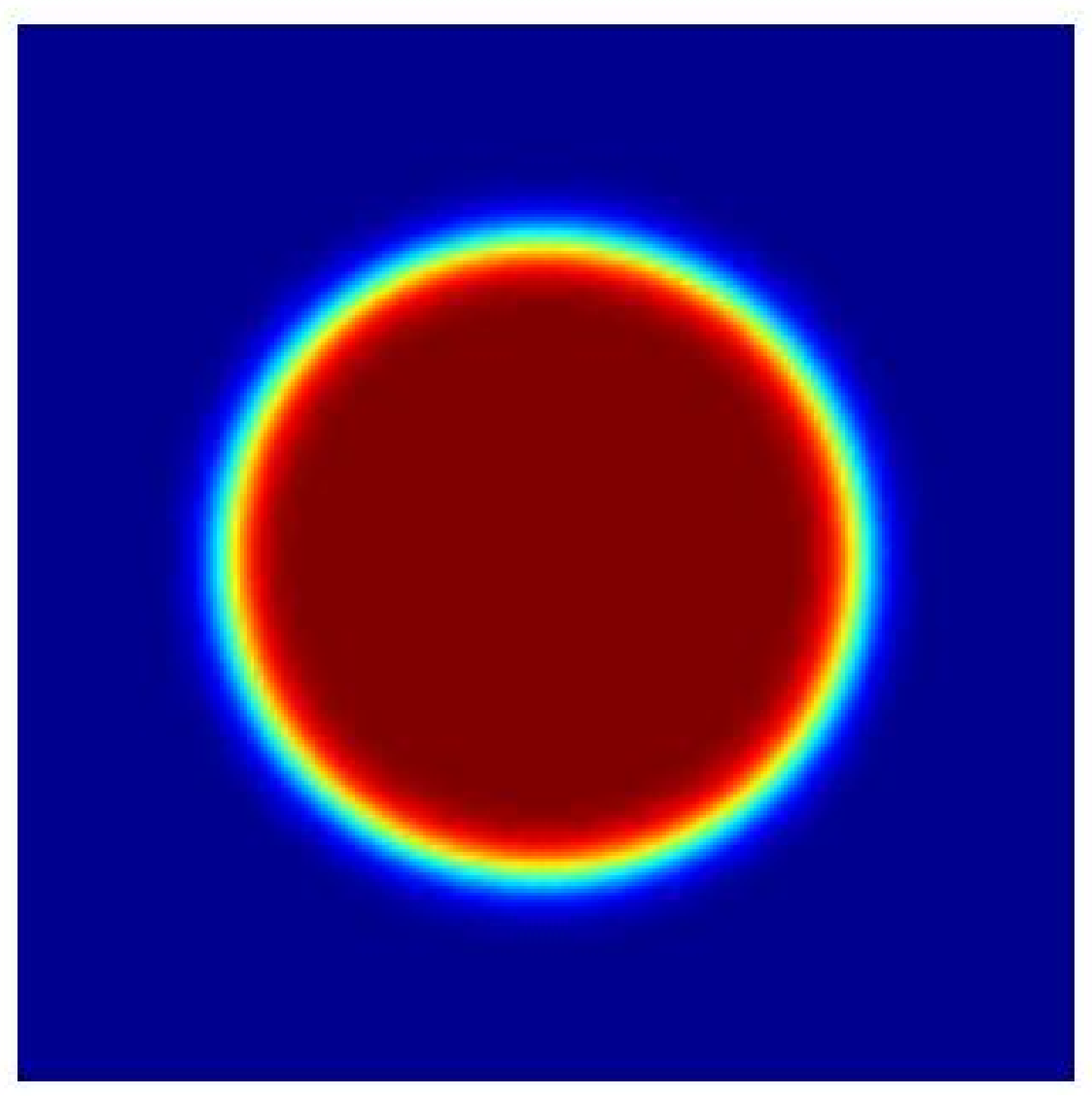}
}
  \subfigure[$\; t = 3000 $]{\label{fig:L2zt3k}
    \centering
 \includegraphics[width=1.5in, height=1.2in, angle=0]{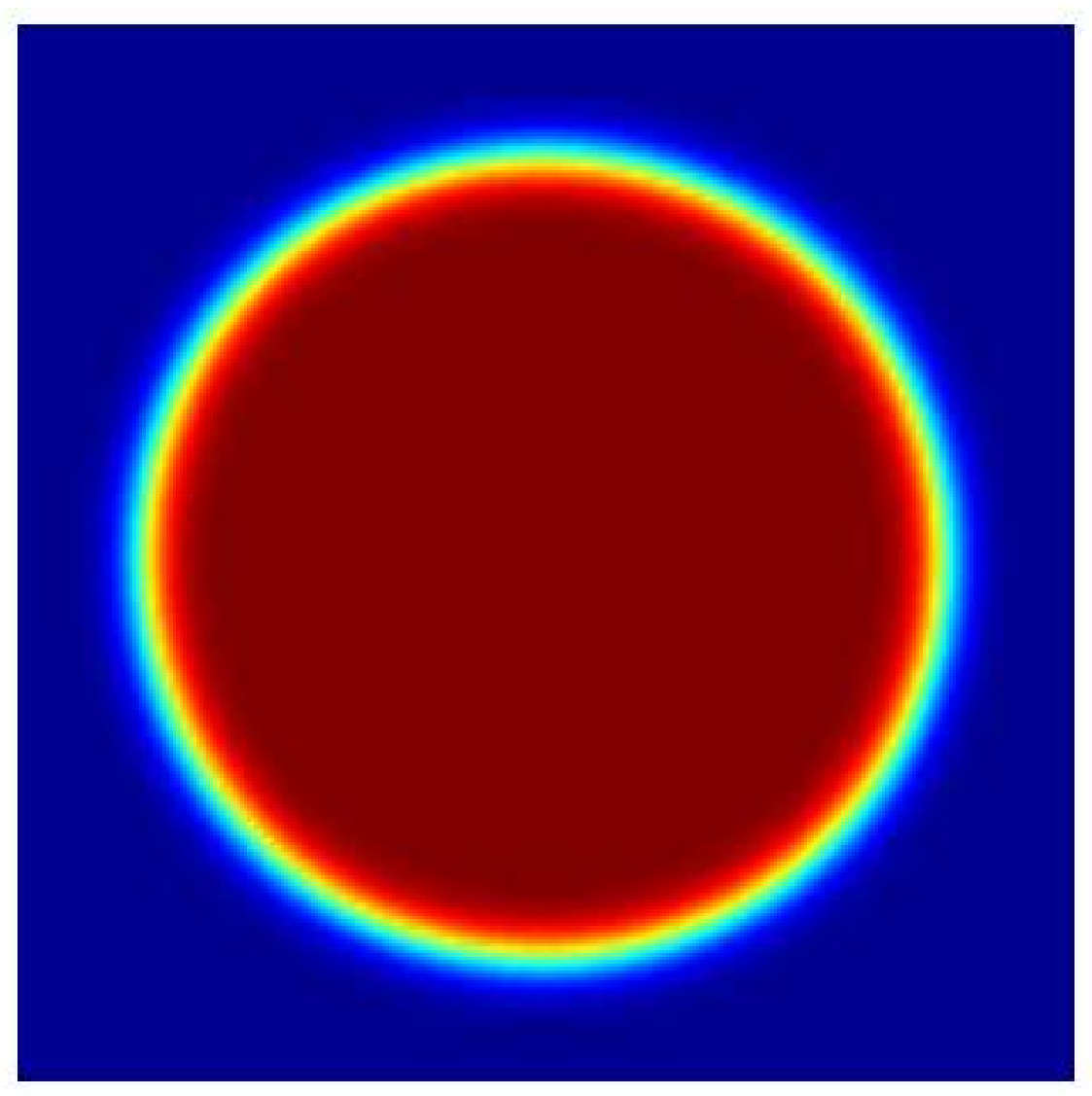}
}
\subfigure[$\; t = 0 $]{\label{fig:L2gt0}
    \centering
 \includegraphics[width=1.5in, height=1.2in, angle=0]{L2zt0.eps}
}
\subfigure[$\; t = 300 $]{\label{fig:L2gt300}
    \centering
 \includegraphics[width=1.5in, height=1.2in, angle=0]{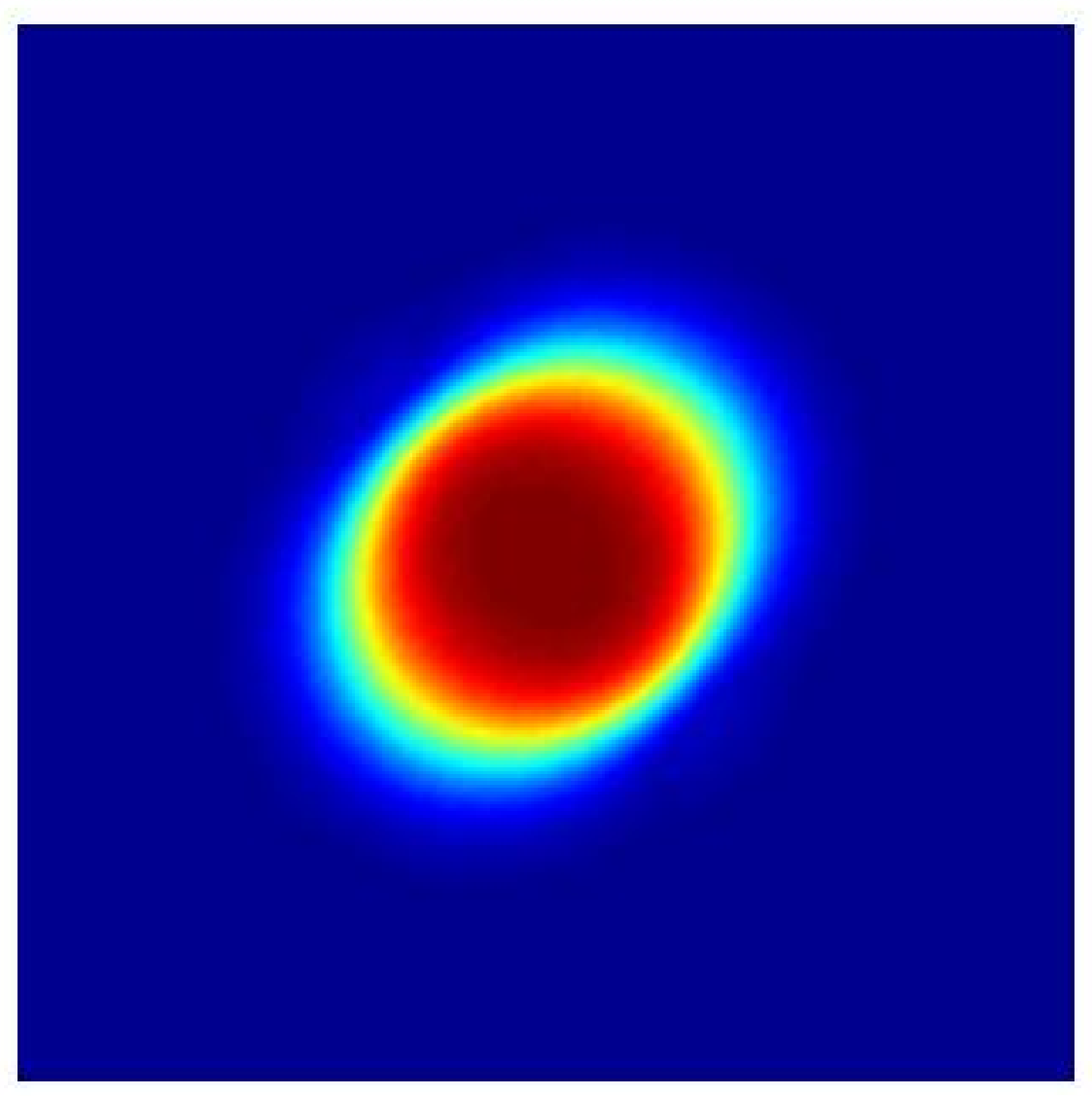}
}
\subfigure[$\; t = 600 $]{\label{fig:L2gt600}
    \centering
 \includegraphics[width=1.5in, height=1.2in, angle=0]{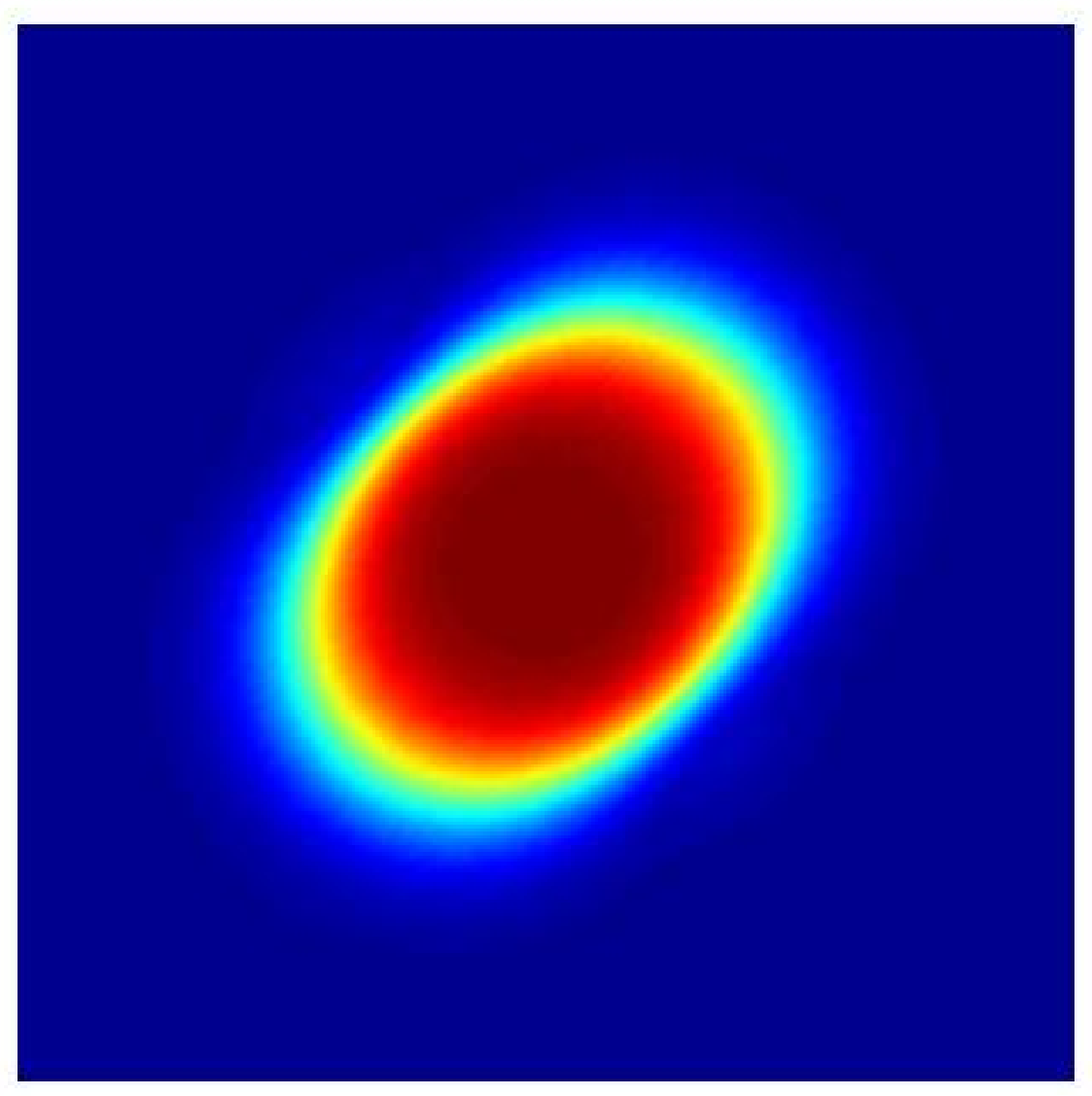}
}
\subfigure[$\; t = 900 $]{\label{fig:L2gt900}
    \centering
 \includegraphics[width=1.5in, height=1.2in, angle=0]{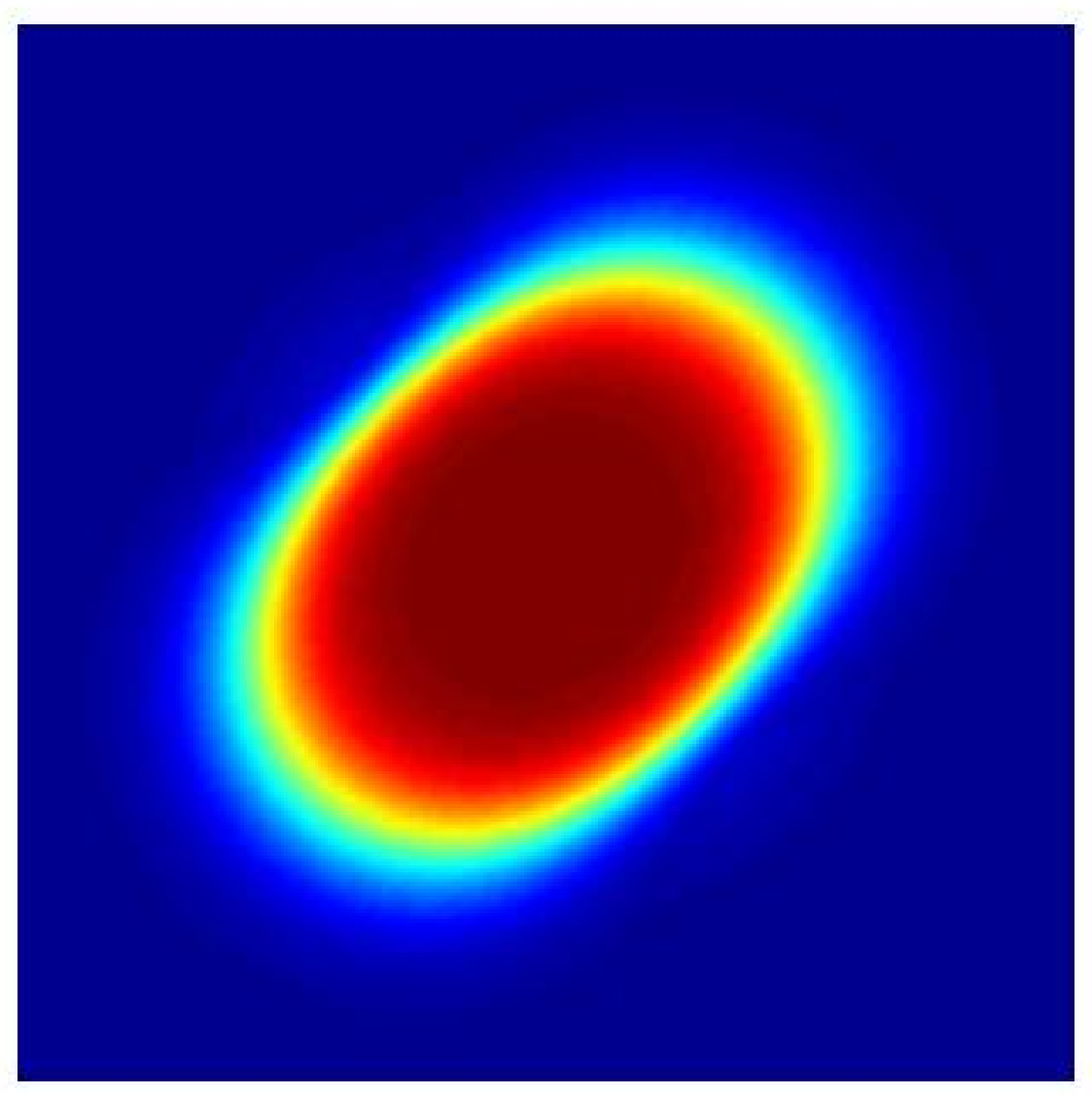}
}
\subfigure[$\; t = 0 $]{\label{fig:L2lt0}
    \centering
 \includegraphics[width=1.5in, height=1.2in, angle=0]{L2zt0.eps}
}
\subfigure[$\; t = 500 $]{\label{fig:L2lt500}
    \centering
 \includegraphics[width=1.5in, height=1.2in, angle=0]{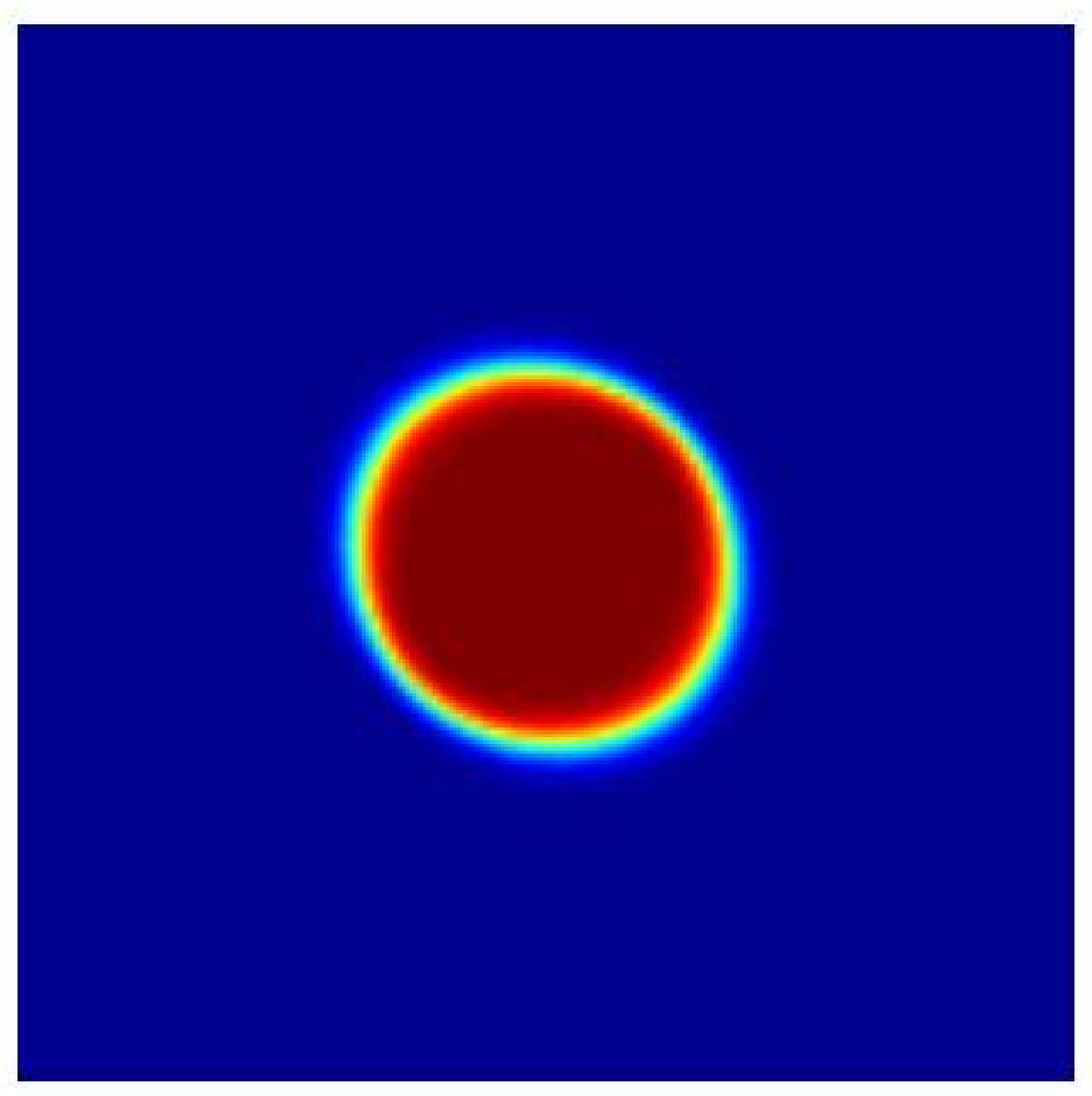}
}
\subfigure[$\; t = 1000 $]{\label{fig:L2lt1k}
    \centering
 \includegraphics[width=1.5in, height=1.2in, angle=0]{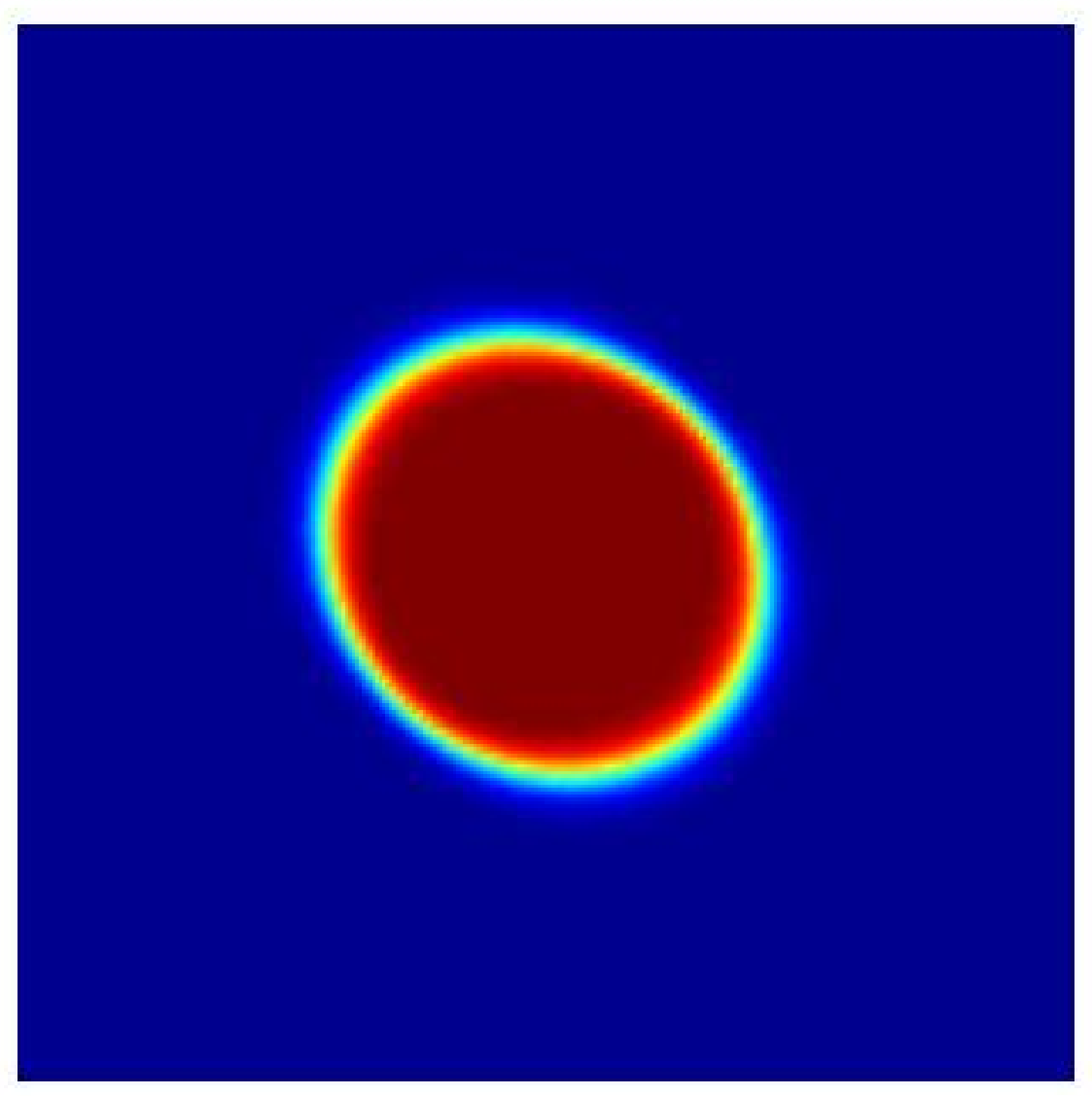}
}
\subfigure[$\; t = 1500 $]{\label{fig:L2lt1500}
    \centering
 \includegraphics[width=1.5in, height=1.2in, angle=0]{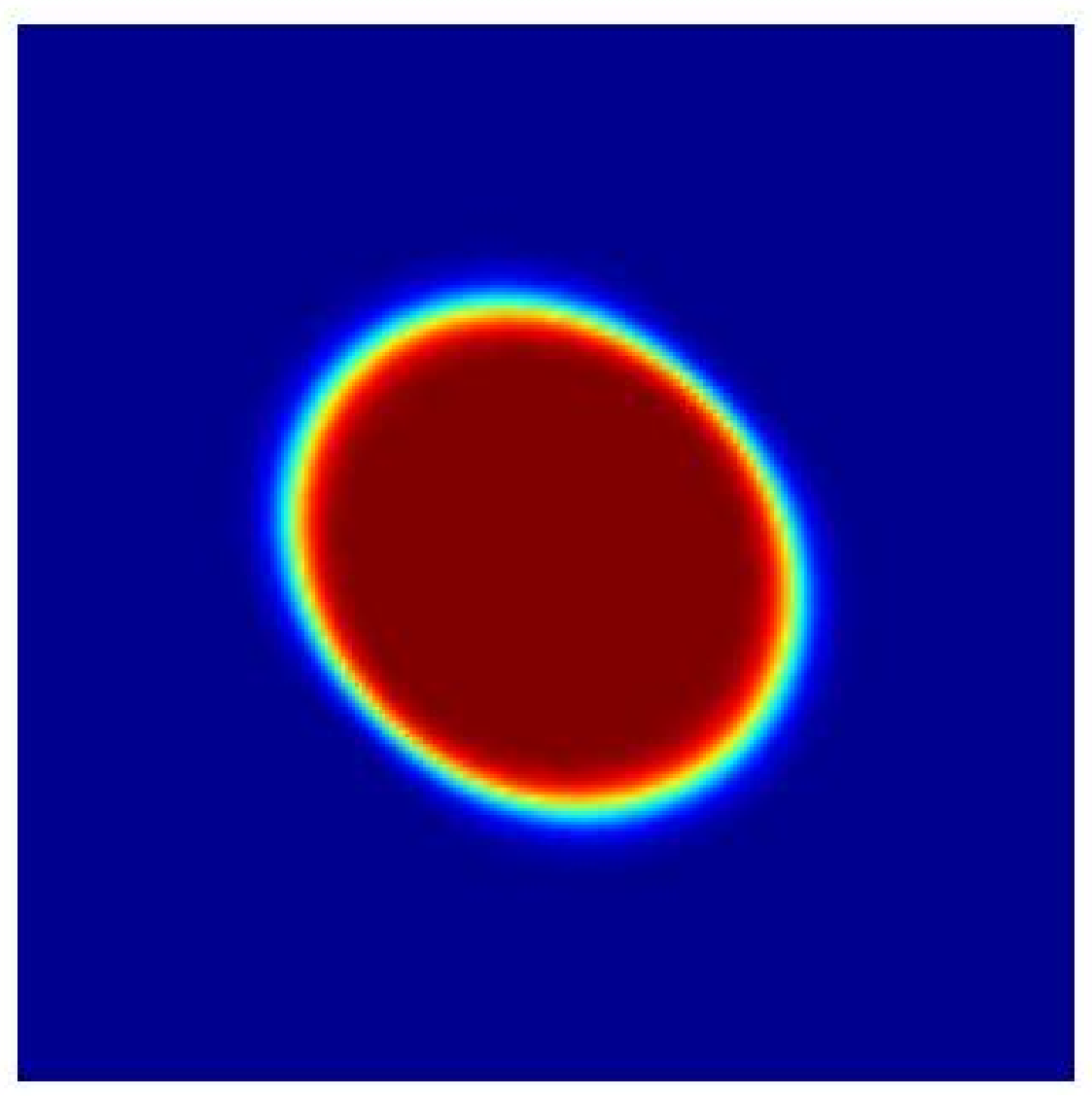}
}
\caption{(Color online) Time evolution of the degree of alignment in the 
relaxation of a nematic droplet. The initially circular droplet develops an anisotropy and 
becomes elliptical, the two-dimensional analogue of the ellipsoidal droplets (tactoids)
seen in three dimensions. Figs.~\ref{fig:L2zt0}, \ref{fig:L2gt0} and \ref{fig:L2lt0} 
shows the circular droplet at time t = 0. The time evolution with $L_2 = 0$ 
of the circular droplet are shown in Figs.~\ref{fig:L2zt1k}, \ref{fig:L2zt2k} 
and \ref{fig:L2zt3k}. The elliptical conformations with $L_2 > 0$ are shown in 
Figs.~\ref{fig:L2gt300}, \ref{fig:L2gt600} and \ref{fig:L2gt900}. These have aspect ratio
1.1191, 1.3046 and 1.5037 respectively.  With $L_2 < 0$ conformations are shown 
in Figs.~\ref{fig:L2lt500}, \ref{fig:L2lt1k} and \ref{fig:L2lt1500}. These have aspect ratio 
1.0122, 1.0718 and 1.1316 respectively. The numerical parameters are A = 0.001, B = -0.5, C = 2.67, $E^{\prime} = 0, 
L_1 = 0.0236, \Gamma = 1.0$, grid size $128 \times 128$ and the droplet radius = $20$. The 
elastic constant $L_2 = 10L_1$ for the $L_2 > 0$ and $L_2 = -L_1$ for $L_2 < 0$.}
\label{fig:drop}
\end{figure}

For droplets larger than the critical radius, the nematic region was
observed to grow in size, finally evolving into the full nematic 
state. For the parameters used, the alignment of the director did 
not change significantly during the evolution of the droplet. 
Figs.~(\ref{fig:drop}) shows three stages in the evolution of a 
single, initially circular droplet, illustrating how the droplet 
shape evolves into an elongated ellipsoidal form. In the absence of elastic
anisotropy $L_2 = 0$, the droplet remains circular. In the presence of 
elastic anisotropy, the droplet orients along the direction of nematic order for
planar anchoring ($L_2 > 0$) and perpendicular to  it for homeotropic anchoring ($L_2 < 0$).

In our numerics, we quantify this geometrical change through 
measurements of the change in the aspect ratio. The aspect ratio, 
defined as the ratio  of the major to minor axis of the ellipse, 
are indicated in Figs.~(\ref{fig:drop}). This is calculated by 
extracting a contour at a fixed value of $S$ (say $S_c/2$) 
and fitting it with an ellipse.

To the best of our knowledge, there are no analytical expressions for order parameter
variations for nematic droplets. Thus, our results cannot be compared directly
against analytical theory. However, the results obtained are in qualitative agreement with
previous work on the shape of nematic droplets as a function of anchoring strength 
\cite{prinschoot, herring, chandrasekhar, virga}.

\subsection{Spinodal coarsening}

In the previous section, we studied the dynamics of droplets when the quench 
was to a temperature between the binodal and the spinodal. For a quench below 
the spinodal temperature, the isotropic phase becomes locally unstable to a 
nematic perturbation and the system proceeds to the nematic phase by spinodal 
decomposition. Coherent regions of local nematic order develop in time, with 
a distinct axis of order in each of these domains. Topological defects 
form at the intersections of these differently ordered domains. Coarsening 
proceeds through the annihilation of topological defects, increasing the correlation 
length of local orientational order. 

To study the coarsening kinetics, we start from a random initial 
configuration where we draw the order parameters S and T randomly from 
a Gaussian  random variable with variance proportional to $F_c$, ensuring 
that $0 < T < S$.  We obtain $\cos\theta$ from an uniform distribution 
between -1 and 1,  and choose $\phi$ similarly between 0 and $2\pi$ to 
generate the director, codirector and the joint normal. We then relax the 
system from this initial condition at a temperature below the supercooling 
spinodal temperature. The data presented below is averaged over $100$ different 
initial conditions for a $256\times256$ system with periodic boundary conditions. 

\begin{figure}
  \subfigure[]{\label{fig:rant0}
    \centering
 \includegraphics[width=2.0in, height=1.5in, angle=0]{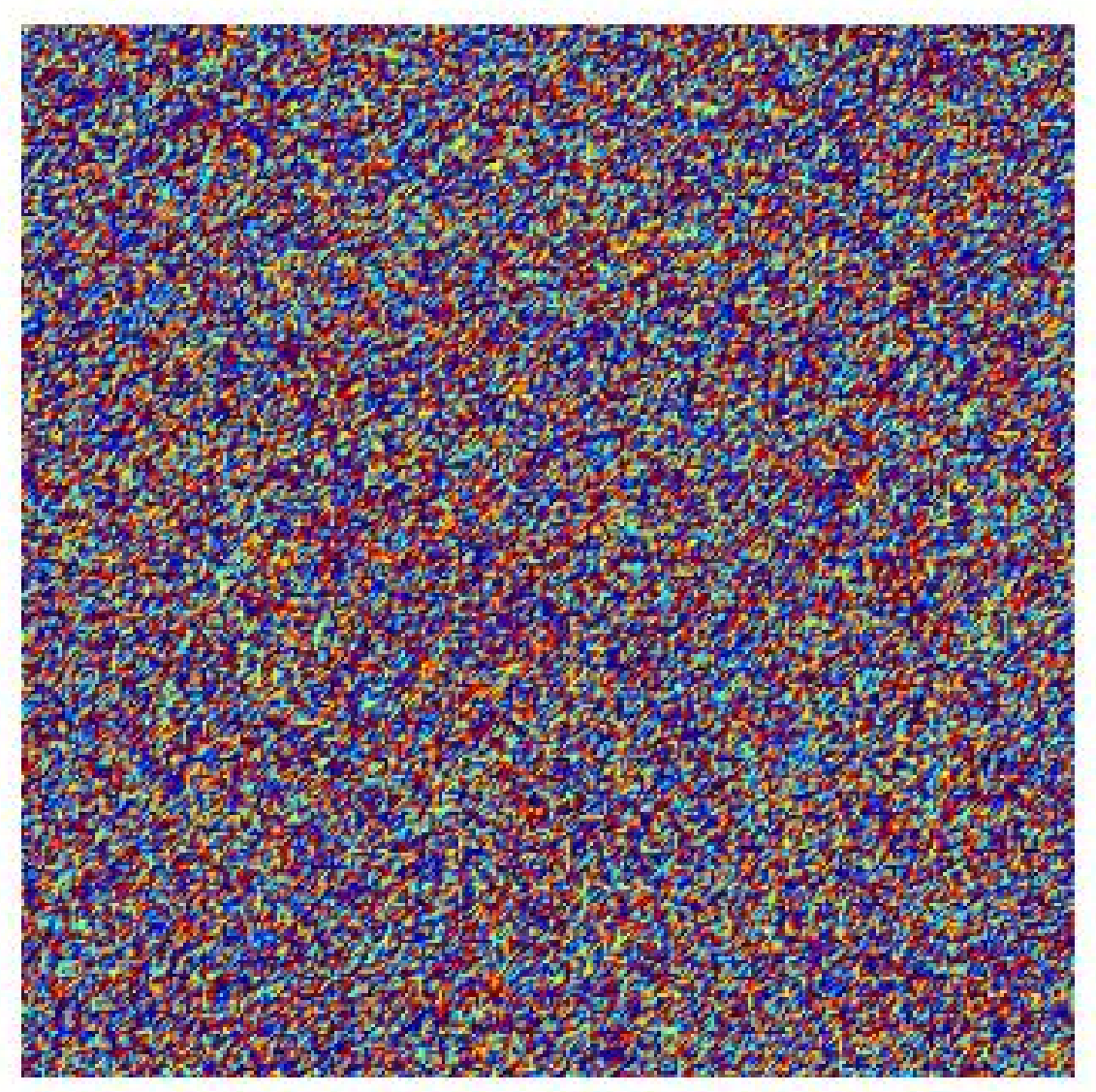}
}
  \subfigure[]{\label{fig:rant1}
    \centering
 \includegraphics[width=2.0in, height=1.5in, angle=0]{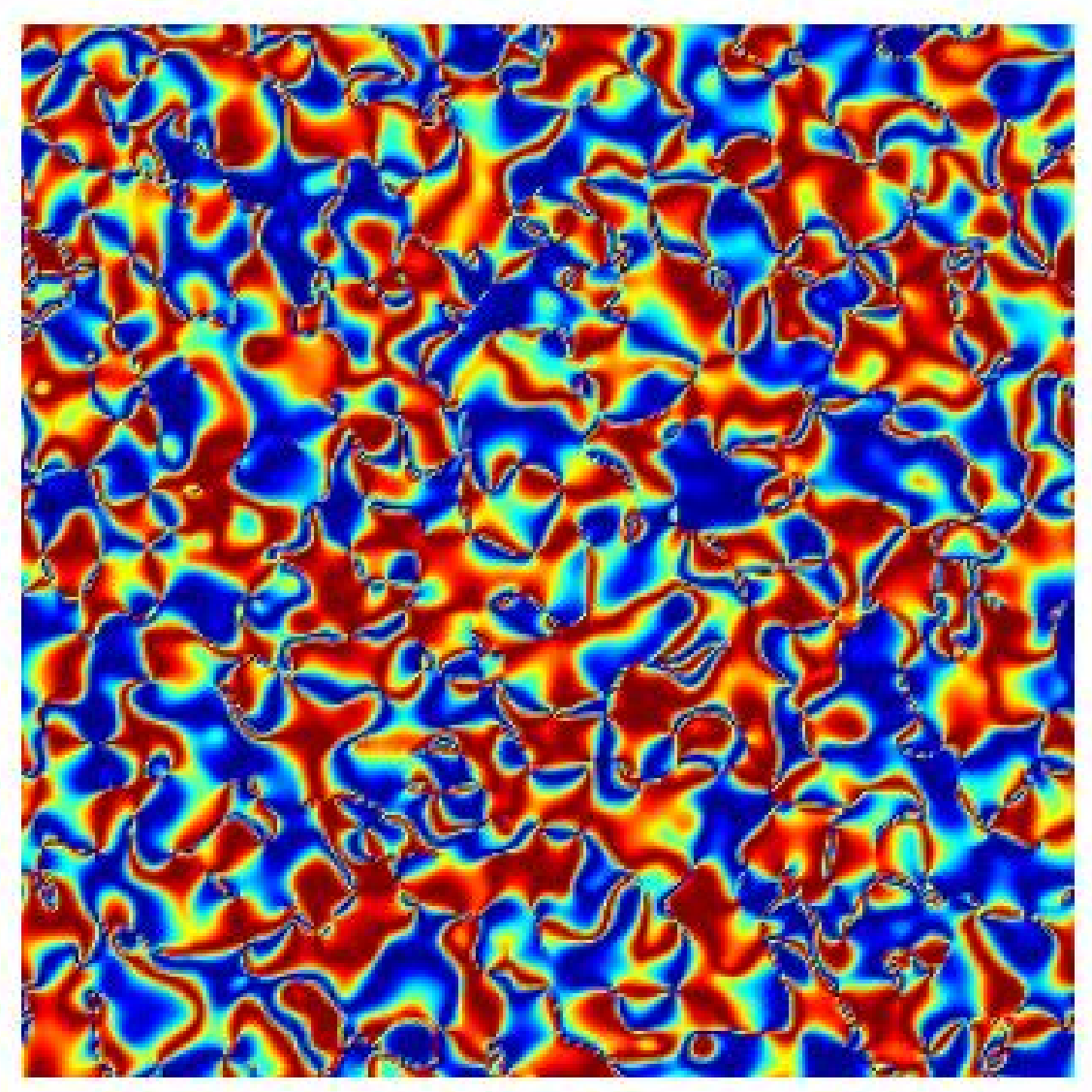}
}     
  \subfigure[]{\label{fig:rant5}
    \centering
 \includegraphics[width=2.0in, height=1.5in, angle=0]{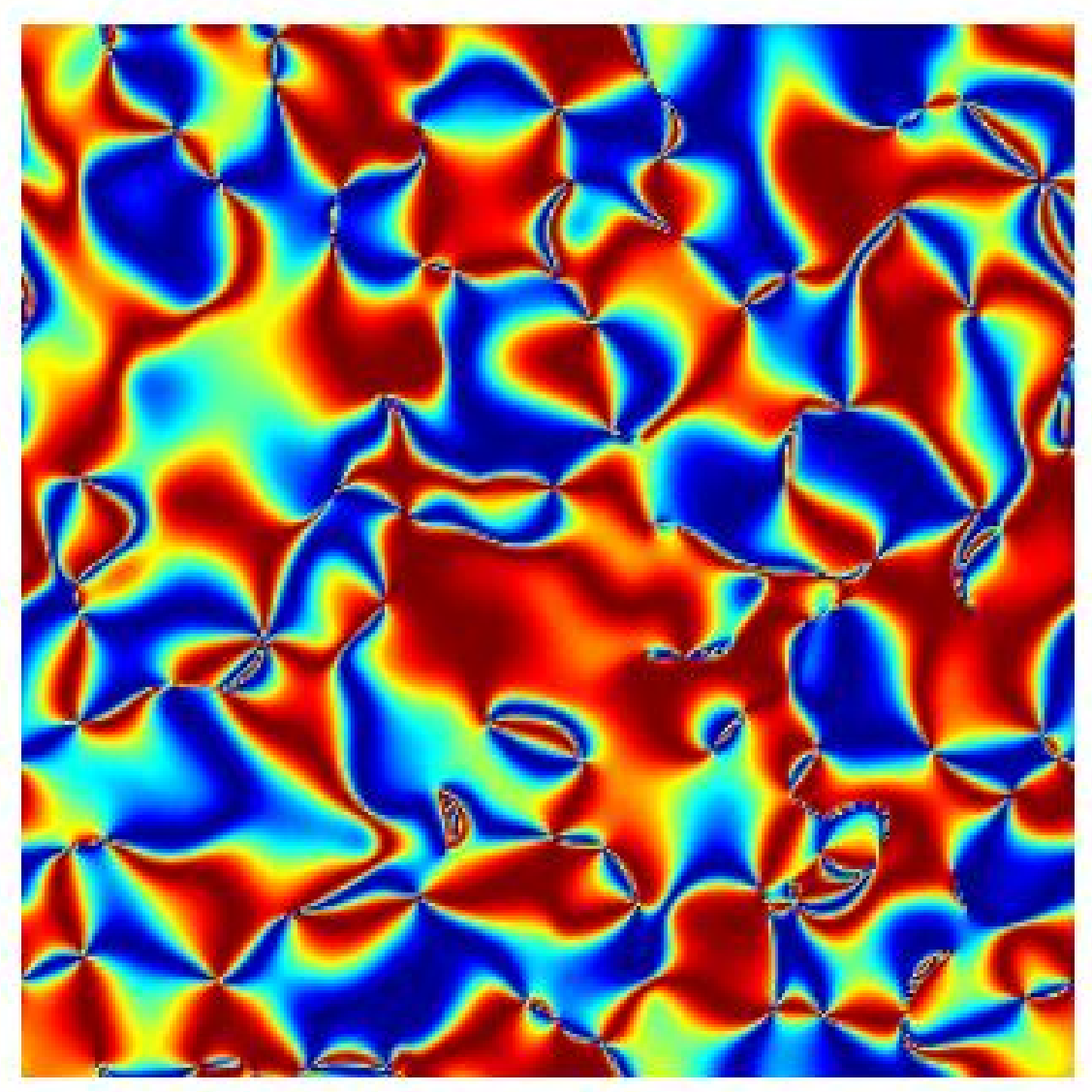}
}     
  \subfigure[]{\label{fig:rant20}
    \centering
 \includegraphics[width=2.0in, height=1.5in, angle=0]{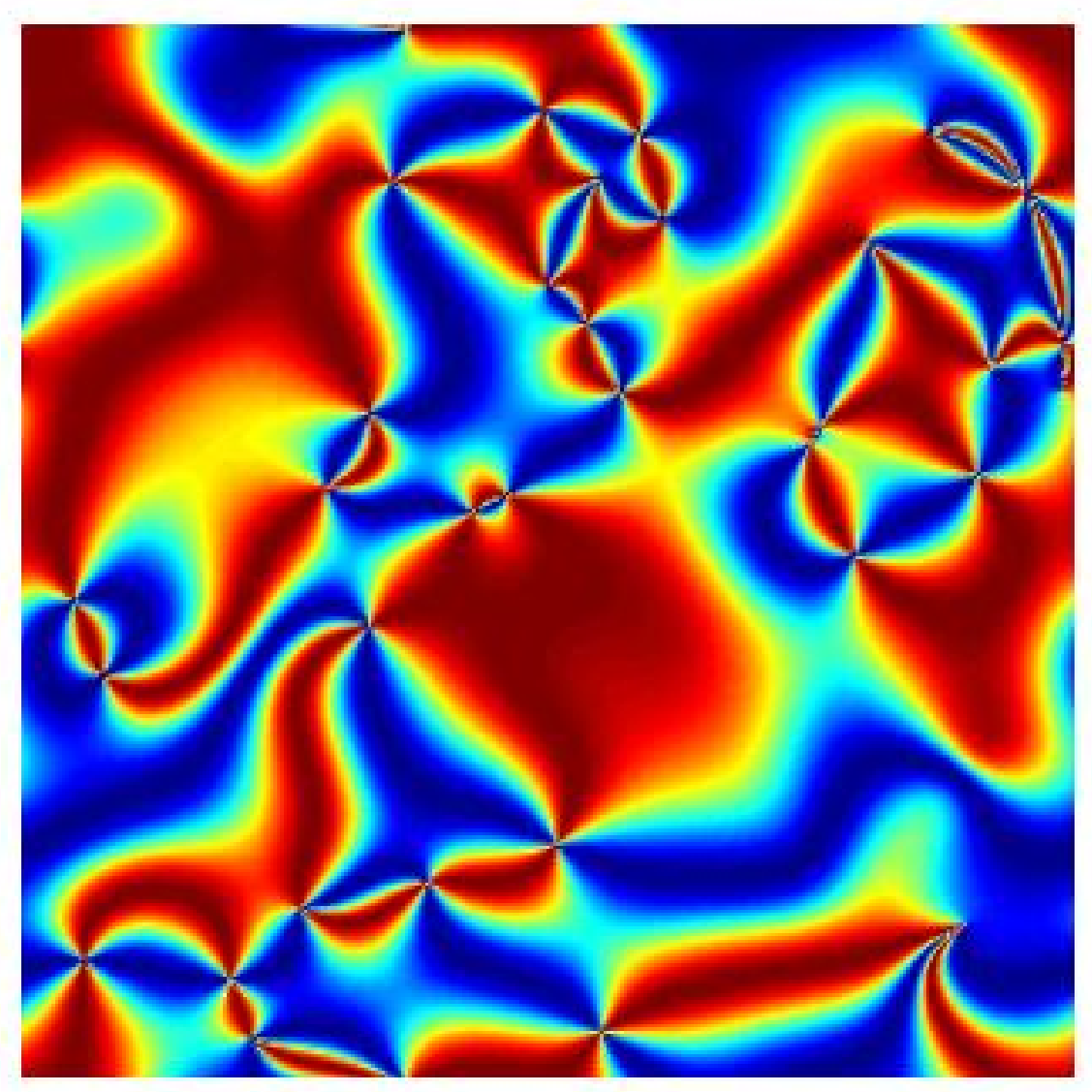}
}     
  \subfigure[]{\label{fig:rant60}
    \centering
 \includegraphics[width=2.0in, height=1.5in, angle=0]{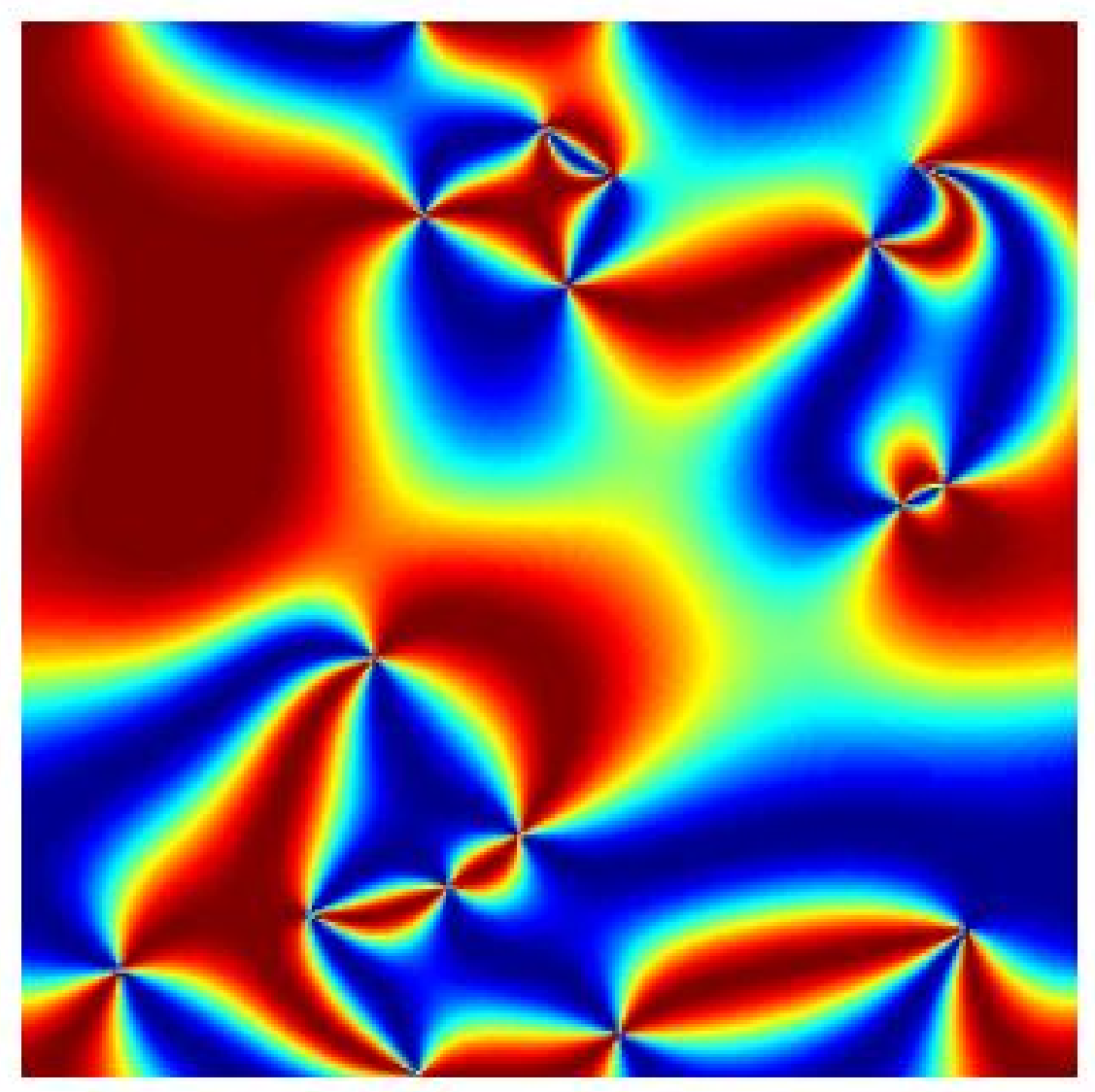}
}     
  \subfigure[]{\label{fig:rant100}
    \centering
 \includegraphics[width=2.0in, height=1.5in, angle=0]{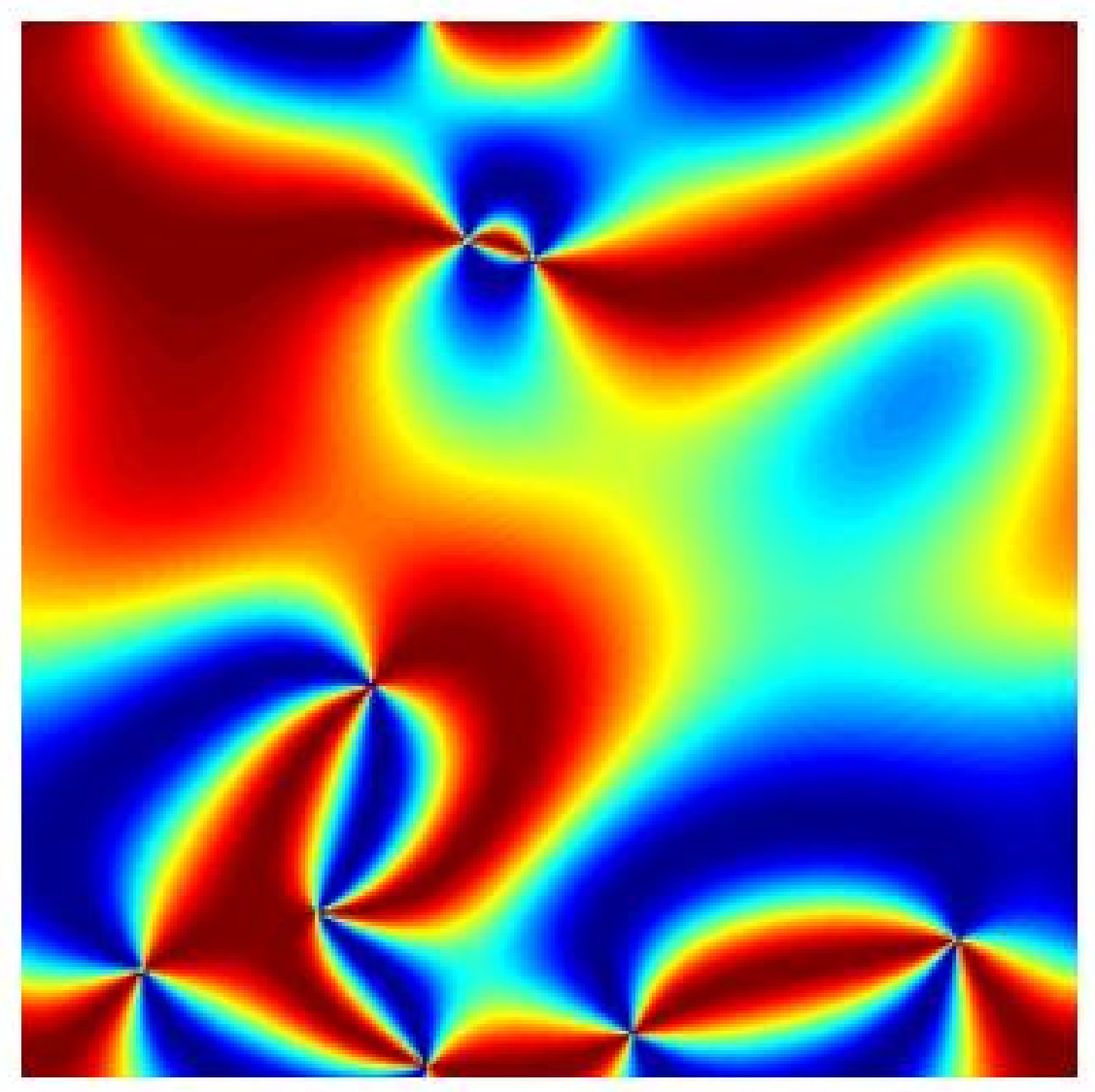}
}     
\caption{(Color online) Schlieren textures in a coarsening nematic. Topological 
defects of both integer charge (top right in (c) and (d)) and half - integer 
charge (throughout) are observed in the simulations. (a) shows the random 
configuration at time t = 0. (b), (c), (d), (e) and (f) show the dynamics 
at different times $t = 10^{3}, 5 \times 10^{3}, 2 \times 10^{4}, 6 \times 10^{4}$ 
and $10^{5}$ respectively. The parameters chosen are A = -0.1, B = -0.5, 
C = 2.67, $E^{\prime} = 0.0, L_{1} = 1.0, L_{2} = 0$ and $\Gamma = 1.0$, 
grid size $256 \times 256$ for $10^{5}$ time steps.}
\label{fig:coarsening}
\end{figure}

\begin{figure}
\subfigure[]{\label{fig:onedefsch}
    \centering
 \includegraphics[width=3.5in, height=2.5in, angle=0]{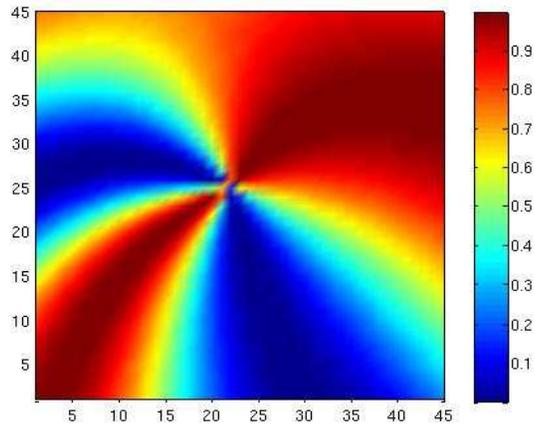}
}
\subfigure[]{\label{fig:onedefect}
    \centering
 \includegraphics[width=3.5in, height=2.5in, angle=0]{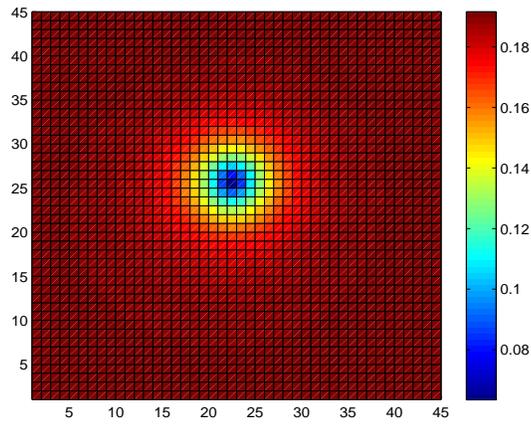}
}
\subfigure[]{\label{fig:defprof}
    \centering
 \includegraphics[width=3.5in, height=2.5in, angle=0]{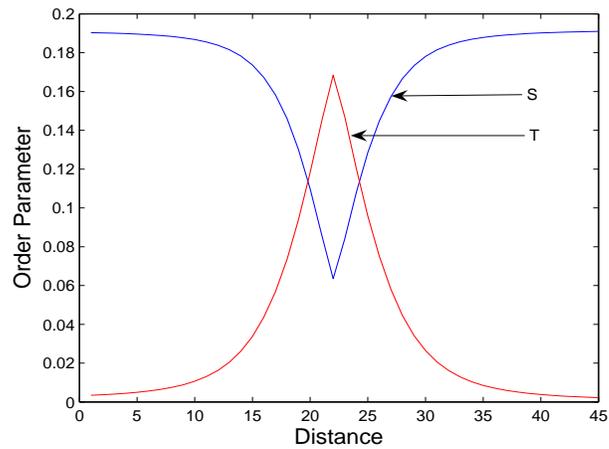}
}
\caption{(Color online) The schlieren texture around a half integer charge defect is shown in (a),
(b) shows the density plot of the uniaxial order parameter, while (c) exhibits the variation of the 
uniaxial and biaxial order parameter along a line passing through the defect core. Note the presence of
strong biaxiality within the defect core as seen previously \cite{schosluck}. The sharpness of the variation 
at the core reflects the coarseness of our discretisation and would be smoothed out by a finer discretisation.}
\label{fig:localdef}
\end{figure}

From the coarsening simulations we obtain the strength of ordering, the biaxiality 
and the director. The director is used to construct the schlieren plots shown
in Figs.~(\ref{fig:coarsening}). These plots are constructed by first projecting 
the director into the $x-y$ plane, finding the angle $\chi$ made by this projection 
with an arbitrary axis (say the x- axis) and then computing $\sin^2(2\chi)$. The 
presence of both integer and half-integer defects is clearly visible in these plots 
as the meeting points of four and two dark brushes, respectively. In the corresponding 
plots for the strength of ordering, the defects are clearly visible as localised 
regions where $S$ rapidly decreases. This is the core region of the topological defect, 
shown in Figs.~(\ref{fig:localdef}). We confirm the surprising finding that there is 
strong biaxial ordering inside the defect core \cite{schosluck}. These results are in perfect qualitative 
agreement with both theoretical predictions and previous numerical results \cite{schosluck}.

\begin{figure}
\begin{center}
\includegraphics[width=3.5in, height=3in, angle=0]{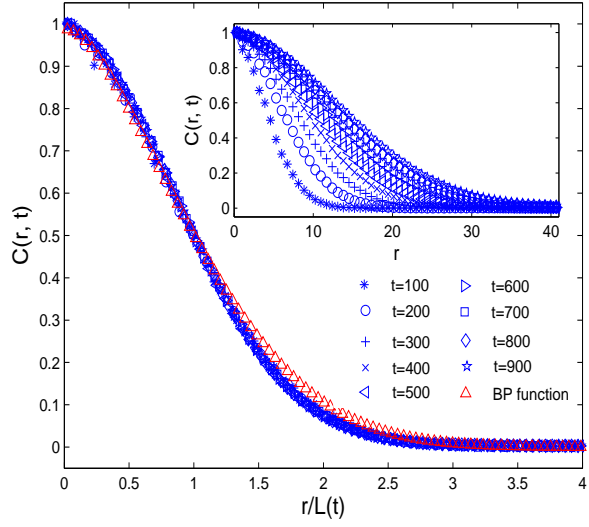}
\caption{(Color online) Data collapse of the direct correlation function $C(r)$ with scaled distance $r/L(t)$ for 
different times. The symbol $\Delta$ depicts the Bray - Puri function \cite{braypuri1} for the O(2) model : 
$f_{BP}(x) = B^2(0.5, 1.5) F[0.5, 0.5, 2; exp(-x^2)] exp(-x^2/2)/\pi$. The inset shows the unscaled correlation 
function at different times. Numerical parameters chosen were $A = -0.1, B = -0.5; C = 2.67; E^\prime = 0, 
L_1 = 1.0, L_2 = 0, \Gamma = 1/20$ on a $256 \times 256$ grid}
\label{fig:dircorr}
\end{center}
\end{figure}

\begin{figure}
\begin{center}
\includegraphics[width=3.5in, height=3in, angle=0]{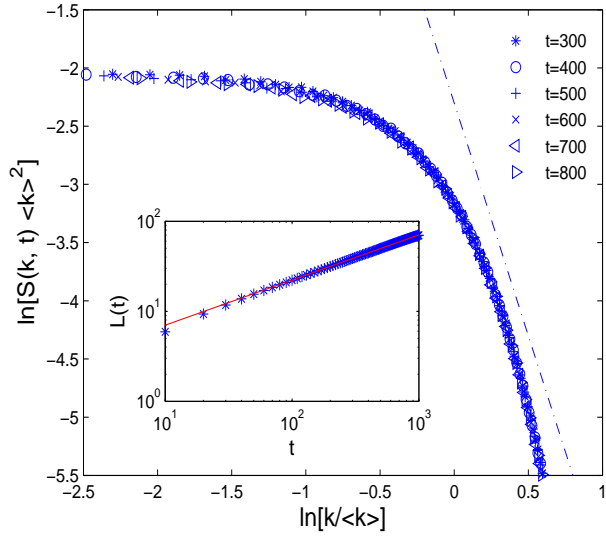}
\caption{(Color online) Data collapse of the structure function $S(k, t)$ at different times. The dash-dot 
line has a slope of $-4$ indicating the validity of generalised Porod's law for O(n) systems. The departure 
from Porod's law at high $k$ is due to the finite core size of the defects as discussed in the text. The 
inset shows the time dependence of the correlation length L(t). The length grows as a power law with an 
exponent of $0.5$. The maximum value of the correlation length is approximately ${1/4}$-th the system size, 
ensuring the absence of finite-size artefacts. Numerical parameters chosen are $A = -0.025, B = -0.5, C = 2.67, 
E^\prime = 0, L_1 = 0.1, L_2 = 0, \Gamma = 1/20$ on a $256 \times 256$ grid. } 
\label{fig:Sqcollapse}
\end{center}
\end{figure}

To make a quantitative comparison with previous work, we compare results for the time-development of correlation functions during coarsening. We calculate 
the real-space correlation function, 
\begin{equation}
C({\bf r}, t) = {\int d^3{\bf x} \;Q_{\alpha\beta}({\bf x}, t) Q_{\beta\alpha}({\bf x + r}, t)\over
                \int{d^3{\bf x} \ Q_{\alpha\beta}({\bf x}, t) Q_{\beta\alpha}({\bf x}, t)}},
\end{equation}
and its Fourier transform,
\begin{equation}
S({\bf k}, t) = {Q_{\alpha\beta}({\bf k}, t) Q_{\beta\alpha}({\bf -k}, t)\over 
                \int{d^3{\bf k} \; Q_{\alpha\beta}({\bf k}, t) Q_{\beta\alpha}({\bf -k}, t)}}.
\end{equation}
Theoretical predictions and analytical work have verified that these correlation 
functions have a scaling form $C(r, t) = {\cal F} [r/L(t)]$, and $S(k,t) = 
L^d(t) {\cal G} [kL(t)]$ \cite{bray}. Here, $C(r, t) = \sum_{|{\bf r}| = r}C({\bf r}, t)$ 
and $S(k, t) = \sum_{|{\bf k}|=k}S({\bf k}, t)$ are angular averages of the correlation 
functions in real and Fourier space respectively. The length $L(t)$ is extracted from the 
real-space correlation function using the implicit condition $C(r=L(t), t) = 1/2$. In 
Fig.~(\ref{fig:dircorr}) we confirm that the real-space correlation function does indeed 
scale as expected. Our numerical data for the scaling function is in close agreement with 
an analytical calculation for the $n=2$ model due to Bray and Puri \cite{braypuri}. 
In Fig.~(\ref{fig:Sqcollapse}) we show the corresponding scaling of the Fourier space 
correlation function. The wavenumber $\langle k\rangle$  is the root of the second moment 
of the S({\bf k}, t) defined by 
\begin{equation}
\langle k\rangle ^2 = \frac{1}{L(t)^2} = \sum_{{\bf k}} k^2 S({\bf k}, t)/\sum_{\bf k}S({\bf k}, t).
\end{equation}
The inset shows the growth of the length scale as a function of time. Theoretically, 
this is expected to grow as a power $L(t)\sim t^{\alpha}$. 
Our estimate for this exponent is $\alpha = 0.5 \pm 0.005$. 

Our results are consistent with both analytical predictions and an earlier 
numerical simulation. The Fourier space correlation function is expected to 
exhibit a short-wavelength scaling $S(k, t)\sim k^{-4}$ known as a 
generalised Porod's law \cite{bray}. We see a clear range of wavenumbers 
where the Porod scaling is obtained. At very short wavelengths, corresponding 
to the size of the defect core, the Porod scaling breaks down. We see evidence 
for this as well, where the very highest wavenumbers in Fig.~(\ref{fig:Sqcollapse}) 
show deviations from the Porod scaling. Our numerical results for spinodal 
decomposition, then, agree both qualitatively and quantitatively with 
theoretical results and previous numerical work \cite{denorlyeom2}. There 
are numerical results reported in the literature are conflicting 
\cite{zapgold2, subhsoumen}. We indicate possible reasons for 
this discrepancy in the following section.

\subsection{Discussion}

For the nematic-isotropic interface, elastic anisotropy ($ L_2 \neq 0$) 
induces biaxiality near the interface. This problem has been studied 
using various approximations by several authors. Sen and Sullivan 
\cite{sensullivan} introduced a symmetry-adapted parametrisation of 
the order parameter which reduces the independent degrees of freedom 
to three. Using this parametrisation Popa-Nita and Sluckin 
\cite{popasluck, popasluckwh} obtained numerical solutions for the 
variation of $S$ and $T$ across the interface. The parametrisation used 
in these calculations is only applicable to the cases of planar or 
homeotropic anchoring at the interface. For a general anchoring condition 
this parametrisation breaks down, as is obvious from the fact that the 
order parameter has, in general, five independent components.

With the present method, the same problem can be studied without any 
approximation, retaining all degrees of freedom of the orientation
tensor. This allows us to obtain the variation of the order parameter 
in situations where the symmetry conditions of Sen and Sullivan do not 
apply. These include curved interfaces and the effect of higher order elasticity.
 
With regard to the problem of nematic droplets a detailed study becomes 
possible, especially in three dimensions. Indeed the only numerical work 
we know of is the Monte Carlo simulation of Cuetos and Dijkstra \cite{cutoedijks}. 
The advantage of the Landau-de Gennes approach for this problem is that 
the free energy of anchoring need not be postulated, (as is done when only the 
director degrees of freedom are retained and a Rapini-Papoular \cite{rapinipap} 
free energy used), but is effectively present in the Landau-de Gennes free 
energy itself. This is the calculational strategy used by Prinsen and van 
der Schoot \cite{prinschoot}. Non-trivial director configurations have been 
proposed for different strengths of bulk-to-surface coupling. Such predictions 
can be verified cleanly with the present method. 

For nematic coarsening in three dimensions, simulations within the Landau-de 
Gennes framework have not been performed so far. The extended nature of topological defects, 
which are lines rather than points in three-dimensional nematics, makes the 
dynamics of the coarsening problem quite different from two dimensions. The 
strength of the Landau-de Gennes approach is particularly clear here. Having 
access to the strength of ordering makes it easy to locate the defect lines 
as tubes of zeros of the ordering strength. This is considerably easier than 
using a Burgers-like circuit integral of the director field to locate defects 
as is done in the work of Zapotocky et al \cite{zapgold2}. 

The dynamics of defect lines has implications in cosmology, where the Kibble 
mechanism \cite{kibble} predicts a scaling relation for the number density of 
defects. Experiments studying the Kibble mechanism in liquid crystals have been 
performed \cite{chuangturrok}. However, we know of no numerical study of the 
Kibble mechanism along the lines proposed here for nematic liquid crystals. 
This provides further impetus for three-dimensional simulations. 

There are two alternatives to the MOL discretisation for tensor order parameter 
descriptions of the nematic phase. These are the cell dynamical scheme of Oono 
and Puri \cite{oonopuri} as implemented by Zapotocky \emph{et al} \cite{zapgold2} 
and Dutta and Roy \cite{subhsoumen}, and the lattice Boltzmann method of Denniston 
\emph{et al} \cite{denorlyeom1}. Our method differs in an important way from both 
these approaches, in that it provides a direct discretisation of the governing 
equations of motions. In cell-dynamical simulations, a local map is constructed 
whose fixed points coincide with the extrema of the local part of the free energy. 
This local map is phenomenological and does not follow from a Landau expansion. 
The advantage of this method is that it produces sharp interfaces, ensuring a 
rapid scale separation between the microscopic interfacial length and the 
macroscopic domain size. In coarsening simulations, this reduces the non-universal 
offset time at which the system enters the scaling regime. 

In comparison, the Landau-de Gennes free energy leads to a less sharp interface. 
Consequently, the structure of the core region of topological defects is different 
in the two methods, being smaller in cell dynamics and larger in the present method. 
We see evidence of this in the dynamic structure factor, Fig.~(\ref{fig:Sqcollapse}), 
where there is a violation of Porod scaling at the highest wavenumbers. As shown in 
Figs.~(\ref{fig:localdef}) the core region of our defect spans several lattice 
spacings in each direction. Thus, as regards scale separation between what are 
microscopic and macroscopic lengths, the cell dynamical scheme fares better. 

On the other hand, when physics at the scale of the interface is important, as in 
the problem of the planar interface and nematic droplets, the MOL discretisation 
offers a clear advantage. Due to the special nature of the cell dynamical Laplacian, 
it is not clear how to generalise the cell dynamical method to include elastic 
anisotropy (corresponding to the gradient term with coefficient $L_2$) or higher 
order terms involving antisymmetric contractions. Due to the phenomenological 
nature of the cell dynamical map, it is difficult to make direct contact with 
experiment, a procedure which is straightforward in the MOL when the 
non-dimensionalisation we outlined above is applied.

In the lattice Boltzmann method, the governing parabolic equations are 
replaced by a hyperbolic superset \cite{succi} through the introduction 
of a distribution function for the tensor order parameter. As with cell 
dynamics, the Landau-de Gennes equations are not solved directly. Rather, 
lattice Boltzmann relies on a temporal scale separation which allows the 
hyperbolic equations to mimic parabolic behaviour. The lattice Boltzmann 
method needs to be carefully formulated \cite{liscarehalid2} to avoid the 
presence of micro-currents and this procedure appears to be non-universal 
\cite{liscarehalid1}. We find no evidence of spurious micro-currents in the 
MOL discretisation presented here. The lattice Boltzmann method includes the 
coupling of order parameter and flow. It is entirely straightforward to extend 
the MOL discretisation to include order 
parameter advection. To include coupling to the fluid momentum, we advocate 
a hybrid strategy, where the order parameter dynamics including advection is 
solved by a MOL discretisation, but the dynamics of the fluid momentum including 
order parameter stresses is solved by the lattice Boltzmann method.

In terms of algorithmic complexity, cell dynamics, lattice Boltzmann and the MOL 
with finite-differences appear to be matched, since both use fairly local information 
to calculate derivatives and involve explicit temporal updates. For $N$ degres of 
freedom the algorithmic complexity of all three algorithms is $O(N)$. On the other 
hand, the storage requirements of the lattice Boltzmann formulation are larger by a 
factor of $6$ to $9$ in two-dimensions, and about $15$ to $19$ in three dimensions. 
On an Intel(R) Core(TM)2 Duo CPU with speed 2.66GHz, our code takes 0.33 seconds for 
one time step once a lattice size of $256\times256$.

\section{Conclusion}

This paper has proposed an efficient numerical scheme based on the method of lines 
for the solving the Landau-de Gennes equations of nematodynamics. The numerical 
results obtained in previous sections are in excellent agreement with analytical 
results where available and consistent with previous numerical data. We expect that 
the method presented here will find broad application in exploring the rich physics 
of the nematic phase of liquid crystals. 

\section{Acknowledgements}

We thank the Indo-French Centre for the Promotion of Advanced Research 
and the DST, India for support.

\bibliography{references}

\end{document}